\documentclass[10pt,twocolumn,floatfix,reprint]{revtex4}

\usepackage{graphicx,amssymb,amsmath}
\usepackage[labelformat=empty]{caption}

\begin{document}

\title{Uncovering the spatial structure of mobility networks}

\author{Thomas Louail$^{1,2}$, Maxime Lenormand$^3$, Miguel
  Picornell$^4$, Oliva Garc\'{i}a Cant\'u$^4$, Ricardo Herranz$^4$,
  Enrique Frias-Martinez$^5$, Jos\'e J. Ramasco$^3$, Marc
  Barthelemy$^{1,6}$\footnote{Correspondence and requests for
    materials should be addressed to MB (Email: marc.barthelemy@cea.fr).}
}
~

\affiliation{$^1$ Institut de Physique Th\'{e}orique, CEA-CNRS (URA 2306), F-91191, 
Gif-sur-Yvette, France}

\affiliation{$^2$G\'eographie-Cit\'es, CNRS-Paris 1-Paris 7 (UMR 8504), 13 rue du
  four, FR-75006 Paris, France}

\affiliation{$^3$IFISC, Instituto de F\'isica Interdisciplinar y Sistemas
Complejos (CSIC-UIB), Campus Universitat de les Illes Balears, E-07122
Palma de Mallorca, Spain}

\affiliation{$^4$Nommon Solutions and Technologies, calle Ca\~nas 8,
  E-28043 Madrid, Spain}

\affiliation{$^5$Telefonica Research, E-28050 Madrid, Spain}

\affiliation{$^6$Centre d'Analyse et de Math\'ematique Sociales, EHESS-CNRS (UMR
8557), 190-198 avenue de France, FR-75013 Paris, France}

\begin{abstract}

  The extraction of a clear and simple footprint of the structure of
  large, weighted and directed networks is a general problem that has
  many applications. An important example is given by
  origin-destination matrices which contain the complete information
  on commuting flows, but are difficult to analyze and compare. We
  propose here a versatile method which extracts a coarse-grained
  signature of mobility networks, under the form of a $2\times 2$
  matrix that separates the flows into four categories. We apply this
  method to origin-destination matrices extracted from mobile phone
  data recorded in thirty-one Spanish cities. We show that these
  cities essentially differ by their proportion of two types of flows:
  integrated (between residential and employment hotspots) and random
  flows, whose importance increases with city size. Finally the method
  allows to determine categories of networks, and in the mobility case
  to classify cities according to their commuting structure.

\end{abstract}

\maketitle

\section*{}
\label{sec:intro}

The increasing availability of pervasive data in various fields has
opened exciting possibilities of renewed quantitative approaches to
many phenomena. This is particularly true for cities and urban systems
for which different devices at different scales produce a very large
amount of data potentially useful to construct a `new science of
cities'~\cite{Batty:2013}.

A new problem we have to solve is then to extract useful information
from these huge datasets. In particular, we are interested in
extracting coarse-grained information and stylized facts that encode
the essence of a phenomenon, and that any reasonable model should
reproduce. Such meso-scale information helps us to understand the
system, to compare different systems, and also to propose models. This
issue is particularly striking in the study of commuting in urban
systems. In transportation research and urban planning, individuals
daily mobility is usually captured in Origin-Destination (OD) matrices
which contain the flows of individuals going from a point to another
(see~\cite{Ortuzar:1994, Weiner:1986}). An OD matrix thus encapsulates
the complete information about individuals flows in a city, at a given
spatial scale and for a specific purpose. It is a large network, and
as such does not provide a clear, synthetic and useful information
about the structure of the mobility in the city. More generally, it is
very difficult to extract high-level, synthetic information from large
networks and methods such as community detection~\cite{Fortunato:2010}
and stochastic block modeling (see for example \cite{Faust:1992} and
\cite{Karrer:2011,Clauset:2014}) were recently proposed. Both these
methods group nodes in clusters according to certain criteria and
nodes in a given cluster have similar properties (for example, in the
stochastic block modeling, nodes in a given group have similar
neighborhood). These methods are very interesting when one wants to
extract meso-scale information from a network, but are unable to
construct expressive categories of links and to propose a
classification of weighted (directed) networks. This is particularly
true in the case of commuting networks in cities, where edges
represent flows of individuals that travel daily from their
residential neighborhood to their main activity area. Several types of
links can be distinguished in these mobility networks, some constitute
the backbone of the city by connecting major residential neighborhoods
to employment centers, while other flows converge from smaller
residential areas to important employment centers, or diverge from
major residential neighborhoods to smaller activity areas. In
addition, the spatial properties of these commuting flows are
fundamental in cities and a relevant method should be able to take
this aspect into account.

There is an important literature in quantitative geography and
transportation research that focuses on the morphological comparison
of cities~\cite{Tsai:2005,Guerois:2008,Schwarz:2010,LeNechet:2012} and
notably on multiple aspects of polycentrism, ranging from schematic
pictures proposed by urban planners and architects~\cite{Bertaud:2003}
to quantitative case studies and contextualized comparisons of
cities~\cite{Salomon:1993,Cattan:2007,Berroir:2011}. So far most
comparisons of large sets of cities have been based on morphological
indicators~\cite{Tsai:2005, Guerois:2008}) --- built-up areas,
residential density, number of sub-centers, etc. --- and aggregated
mobility indicators~\cite{Schwarz:2010,LeNechet:2012} --- motorization
rate, average number of trips per day, energy consumption \emph{per
  capita} per transport mode, etc. ---, and have focused on the
spatial organization of residences and employment centers. But these
previous studies did not propose generic methods to take into account
the spatial structure of commuting trips, which consist of both an
origin and a destination. Such comparisons based on aggregated
indicators thus fail to give an idea of the morphology of the city in
terms of daily commuting flows. We still need some generic methods
that are expressive in a urban context, and that could constitute the
quantitative equivalent of the schematic pictures of city forms that
have been pictured for long by urban planners~\cite{Bertaud:2003}.

In this paper we propose a simple and versatile method designed to
compare the structure of large, weighted and directed networks. In the
next section we describe this method in detail. The guiding idea is
that a simple and clear picture can be provided by considering the
distribution of flows between different types of nodes. We then apply
the method to commuting (journey to work) OD matrices of thirty-one
cities extracted from a large mobile phone dataset. We discuss the
urban spatial patterns that our method reveals, and we compare these
patterns observed in empirical data to those obtained with a
reasonable null model that generates random commuting
networks. Finally the method allows determining categories of networks
with respect to their structure, and here to classify cities according
to their commuting structure. This classification highlights a clear
relation between commuting structure and city size.

\section*{Results}
\label{sec:results}

\subsection*{Extracting coarse-grained information from OD matrices}

For the sake of clarity, we will use here the language of OD matrices,
but the method could easily be applied to any weighted and directed
network from which we want to extract high-level information.

We assume that for a given city, we have the $n\times n$ matrix
$F_{ij}$ where $n$ is the number of spatial units that compose the
city at the spatial aggregation level considered (for example a grid
composed of square cells of size $a$, see Methods). This OD matrix
$F_{ij}$ represents the number of individuals living in the location
$i$ and commuting to the location $j$ where they have their main,
regular activity (work or school for most people). By convention, when
computing the numbers of inhabitants and workers in each cell we do
not consider the diagonal of the OD matrix. This means that we omit
the individuals who live and work in the same cell (considered as
`immobile' at this spatial scale).

In order to extract a simple signature of the OD matrix, we proceed in
two steps. We first extract both the residential and the work
locations with a large density --- the so called `hotspots'
(see~\cite{Louail:2014}). The number of residents of cell $i$ is
given by $\sum_{j\neq i}F_{ij}$ and its number of workers is given by
$\sum_{j\neq i} F_{ji}$. The hotspots then correspond to local maxima
of these quantities. It is important to note that the method is
general, and does not depend on how we determine these hotspots.

Once we have determined the cells that are the residential and the
work hotspots (some cells can possibly be both), we proceed to the
second and main step of the method. We reorder the rows and columns of
the OD matrix in order to separate hotspots from non-hotspots. We put
the $m$ residential hotspots on the top lines, and do the same for
columns by putting the $p$ work hotspots on the left columns. The OD
matrix then becomes a 4-quadrants matrix where the flows $F_{ij}$ are
spatially positioned in the matrix with respect to their nature: on
the top left the individuals that live in hotspots and work in
hotspots; at the top right the individuals that live in hotspots and
do not work in hotspots; at the bottom left individuals that do not
live in hotspots but work in hotspots; and finally in the bottom right
corner the individuals that neither live or work in hotspots. For each
quadrant we sum the number of commuters and normalize it by the total
number of commuters in the OD matrix, which gives the proportion of
individuals in each of the four categories of flows. In other words,
for a given city, we reduce the OD matrix to a $2\times 2$ matrix
\begin{equation}
\Lambda=\begin{pmatrix} I&D\\ C&R
\end{pmatrix}
\end{equation}
where
%\begin{itemize}
%\item 

$$I = \sum_{i \in 1..m, j \in 1..p} F_{ij} / \sum_{i,j \in 1..n}
F_{ij}$$ is the proportion of \textbf{I}ntegrated flows that go from
residential hotspots to work hotspots;

$$C = \sum_{i \in m+1..n, j \in 1..p} F_{ij} / \sum_{i,j \in
  1..n} F_{ij}$$ is the proportion of \textbf{C}onvergent flows that go
from random residential places to work hotspots;

$$D = \sum_{i \in 1..m, j \in p+1..n} F_{ij} / \sum_{i,j \in
  1..n} F_{ij}$$ is the proportion of \textbf{D}ivergent flows that go
from residential hotspots to random activity places;

$$R = \sum_{i \in m+1..n, j \in p+1..n} F_{ij} / \sum_{i,j \in
  1..n} F_{ij}$$ is the proportion of \textbf{R}andom flows that occur
`at random' in the city, i.e. that are going from and to places that
are not hotspots.

\begin{figure}
  \centering
  \includegraphics{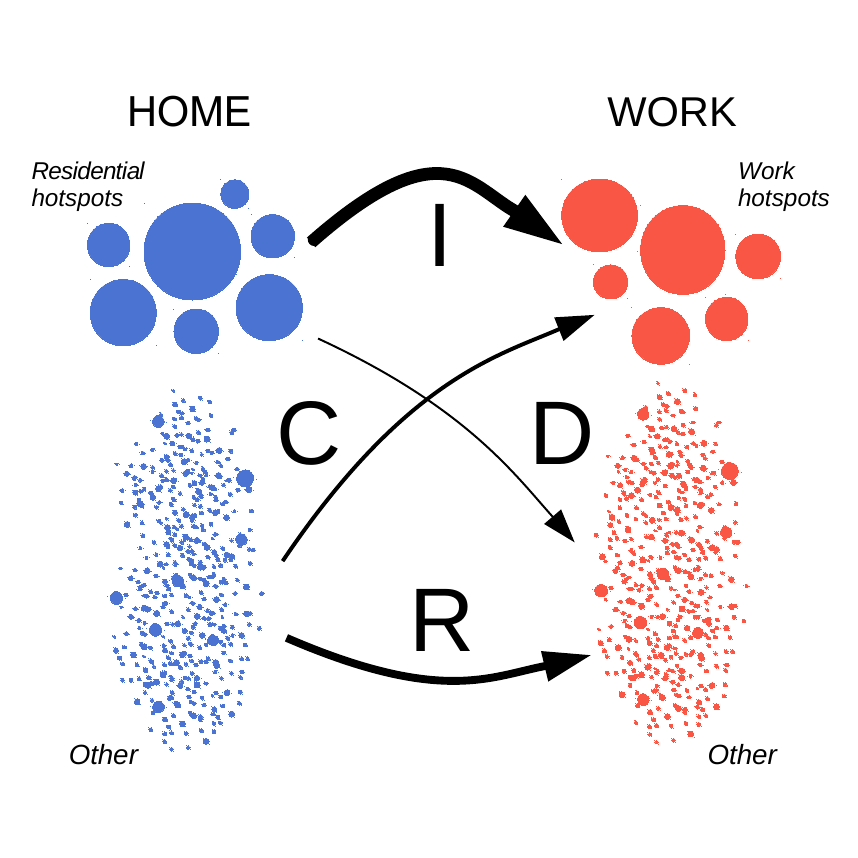}
  \caption{{\bf Illustration of the ICDR method.} The method
    decomposes the commuting flows in the city in four categories: the
    Integrated flows ($I$) from hotspot to hotspot, the Convergent
    flows ($C$) to hotspots, the divergent flows ($D$) originating at
    hotspots and finally the random flows ($R$), which are neither
    starting nor ending at hotspots. For each city with its
    origin-destination matrix, we can compute the importance of each
    commuting flow category and get a simple picture of the mobility
    structure in the city.}
  \label{fig:ICDR}
\end{figure}

By construction,  we have $I, C, D, R \in [0;1]$ and $I + C + D + R
=1$. This matrix $\Lambda$ is thus a very simple footprint of
the OD matrix that gives an expressive picture of the structure of
commuting in the city, as illustrated by Fig.~\ref{fig:ICDR}.

%%% Figure 1 : Method ICDR

\subsection*{Comparison of the mobility networks of thirty-one cities}

\subsubsection*{Commuting data}
Large scale individual mobility networks are nowadays extracted from
pervasive geolocated data, such as mobile phone, GPS, public transport
cards or social apps
data~\cite{Schneider:2013,Roth:2011,Noulas:2012,Wu:2014,Zhong:2014}. In
particular, if an individual's mobile phone geolocated activity is
available during a sufficiently long period of time, it is possible
--- under certain regularity conditions --- to infer the most likely
locations of her home and her workplace, and by aggregation to
construct OD
matrices~\cite{Kung:2013,Jiang:2013,LeNormand:2014}. Several
parameters however impact the construction of OD matrices such as the
nature of the data source (survey or user-generated geolocated data),
or the spatial scale at which the OD matrix is built which can be
dictated by administrative units (divisions in wards, counties,
municipalities, etc.) or by technical reasons such as the density of
antennas in the case of mobile phone data. Given this variety of data
collection protocols, it is thus particularly remarkable that when
considering the commuting flows at a city scale, various sources of
pervasive data provide a very similar mobility information when
compared to the OD matrices built from
surveys~\cite{LeNormand:2014}. This result needs to be confirmed for
other cities and countries, but it already opens the door to a
systematic use of pervasive, geolocated data as a relevant substitute
to traditional transport surveys.

In the following we apply our ICDR method to OD matrices that have
been extracted from mobile phone records in thirty-one Spanish urban
areas during a five weeks period (see the Methods section for details
on the dataset and the calculation of the OD matrices).

\subsubsection*{Hotspots}

As described above, the first step consists in determining
hotspots. Several possible methods have been proposed in the
literature~\cite{Berroir:2011,McMillen:2001,Griffith:1981}, and we use
here a parameter-free method based on the Lorenz curve of the
densities that we have proposed in a recent study
(see~\cite{Louail:2014} and the Methods section). Once we have
determined both the origin (residential) hotspots and the destination
(work) hotspots in each city, we first observe how their number scale
with the population size of the city. Both these numbers for
residential and employment hotspots scale sublinearly with the
population size (see Supplementary Figures 4 and 5). The number of
work hotspots grows significantly slower than the number of
residential hotspots, showing that residential areas are (i) more
dispersed in the city, and (ii) are more numerous than activity
centers, as intuitively expected (see Supplementary Figure 6 that
displays the locations of Home and Work hotspots in four cities that
exhibit different spatial organizations). We also note here that the
sublinear scaling of the work hotspots confirms previous results
obtained with a totally different dataset (the number of employment
centers in US cities) \cite{Louf:2013}.

\subsubsection*{ICDR values}

\begin{figure*}
  \centering
   \includegraphics[width=.8\linewidth]{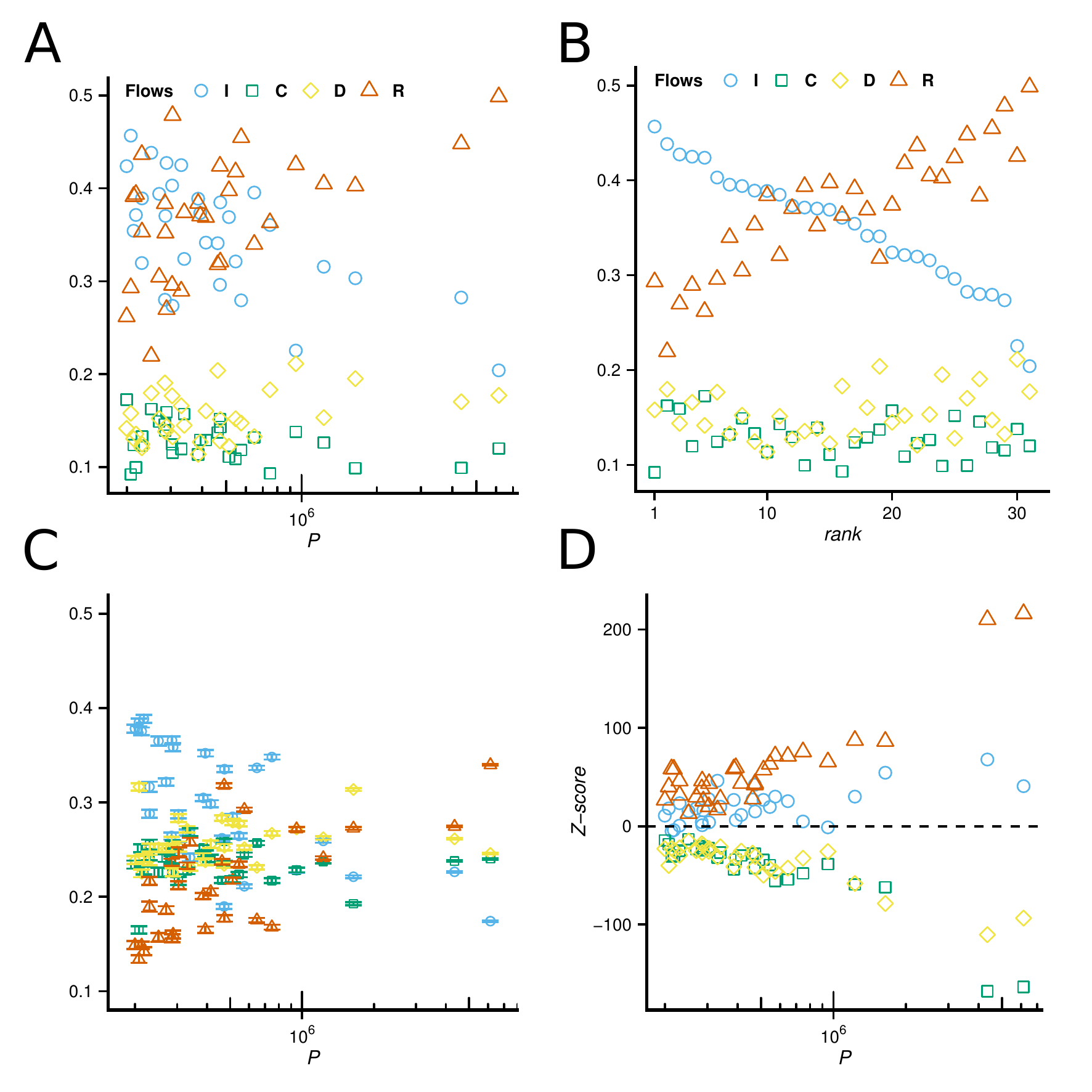} 
   \caption{{\bf Results for 31 Spanish cities.} (a) $I$ (integrated),
     $C$ (convergent), $D$ (divergent) and $R$ (random) values versus
     population size for $31$ Spanish urban areas. (b) Same ICDR
     values as in (a) but sorted by decreasing order of $I$ (note that
     by definition, we have for each city $I+C+D+R=1$). It is
     remarkable that $I$ and $R$ dominate and seem almost sufficient
     to distinguish cities, while $C$ and $D$ are almost constant
     whatever the city size (see Supplementary Figure 7 for the values
     obtained with another size $a$ of grid cells). (c) $I,C,D,R$
     average values and standard deviations obtained for $100$
     replications of a null model, where the inflow and outflow at
     each node are kept constant while flows are randomly distributed
     at random between nodes. (d) Z-scores obtained by comparing the
     empirical data and the values returned by the null model. Large
     values of Z-scores show that the actual commuting networks cannot
     be considered as resulting from connecting the nodes at
     random. The $I$, $C$, $D$ and $R$ values of a specific city are
     then a signature of its structure.}
  \label{fig:ICDR-values}
\end{figure*}

We now apply the second part of the method in order to calculate the
$I$,$C$,$D$ and $R$ values for each OD matrix. For the 31 Spanish
urban areas under study (see Supplementary Figure 1), we obtain the
values shown in Fig.~\ref{fig:ICDR-values}. In
Fig.~\ref{fig:ICDR-values}(a), we plot these values versus the
population size of these cities. For this sample of cities we see that
globally the proportion $I$ of individuals that commute from hotspot
to hotspot decreases as the population size increases, while the
proportion $R$ of `random' flows increases and the proportions $C$ and
$D$ of convergent and divergent flows seem surprisingly constant
whatever the city size. In Fig.~\ref{fig:ICDR-values}(b) we plot the
same values but sorted by decreasing values of $I$ which shows clearly
that the $I$ and $R$ values are the relevant parameters for
distinguishing cities from each other.

% Figure 2 : ICDR values

We also notice that the values obtained for another spatial scale of
data aggregation confirm this trend (see Supplementary Figure 7). The
decay of $I$ flows ('integrated') flows in favor of $R$ flows
('random') when $P$ increases shows that the population growth among
Spanish cities goes with a decentralization of both activity places
and residences. As cities get bigger, their numbers of residential and
employment hotspots grow (sublinearly), but these hotspots catch a
smaller part of the commuting flows.

\subsubsection*{A null model} 

In order to evaluate to what extent the $ICDR$ signatures of cities
are characteristic of their commuting structure, we compare these
values to the ones returned by a null model of commuting flows. For
each city we generate random OD matrices of the same size than the
reference OD matrix but with random flows of individuals that preserve
the in- and out- degree of each node (see Supplementary Note
5). Fig~\ref{fig:ICDR-values}(c) shows the average values and standard
deviation obtained for 100 replications. On
Fig~\ref{fig:ICDR-values}(d) we plot the Z-scores of the $I$, $C$,
$D$, and $R$ values of each city when compared to the values $I^*$,
$C^*$, $D^*$ and $R^*$ returned by our null model (e.g. for the
quantity $R$ of the city $i$, the Z-score is given by $Z(R_i)=(R_i -
\overline{R_i^*}) / \sigma(R_i^*)$).  Essentially, we observe that the
Z-scores of $I$ and $R$ are positive and large, while those of $C$ and
$D$ are negative (and large in absolute value). Also as cities grow,
the Z-scores of $I$ and $R$ increase while those of $C$ and $D$
decrease. These results demonstrate that the larger a city, and the
less random it appears. This is in contrast with the naive expectation
that the larger a city and the more disordered is the structure of
individuals' mobility. For large cities, there seems to be a commuting
backbone which cannot result from purely random movements of
individuals. This backbone is the footprint of the city's structure
and history, and probably results from strong constraints and
efficiency considerations.

\subsubsection*{Robustness} 

Since our method first requires to determine origin and destination
hotspots, one could argue that the interpretation of the $I$,$C$,$D$
and $R$ values will crucially depend on the particular method chosen
to define these hotspots. The identification of hotspots is a problem
that has been broadly discussed in urban economics (see Supplementary
Note 3). Roughly speaking, starting from a spatial distribution of
densities, the goal is to identify the local maxima and amounts to
choose a threshold $\rho^*$ for the density $\rho$ of individuals: a
cell $i$ is a hotspot if the local density of people is such that
$\rho(i) >\rho^*$. In order to test the impact of the choice of a
particular threshold $\rho^*$ on the resulting $ICDR$ values, we
measure the sensitivity of these values as a function of the density
threshold $\rho^*$ (see Supplementary Figure 8). As expected, the
lower the density threshold, the larger the number of origin and
destination hotspots, and consequently the larger $I$ and the smaller
$R$. In contrast, changing the density threshold has little impact on
the $C$ and $D$ terms. More importantly, the conclusions drawn from
the comparison of the $ICDR$ values across cities remain the same:
whatever the density threshold $\rho^*$ chosen to define
residential/employment hotspots, we still observe the same qualitative
behavior: a decay of integrated flows ($I$) in favor of `random' flows
($R$), when the population size increases.

\subsubsection*{Distance of each type of flows}

\begin{figure*}
  \centering
  \includegraphics[width=.8\linewidth]{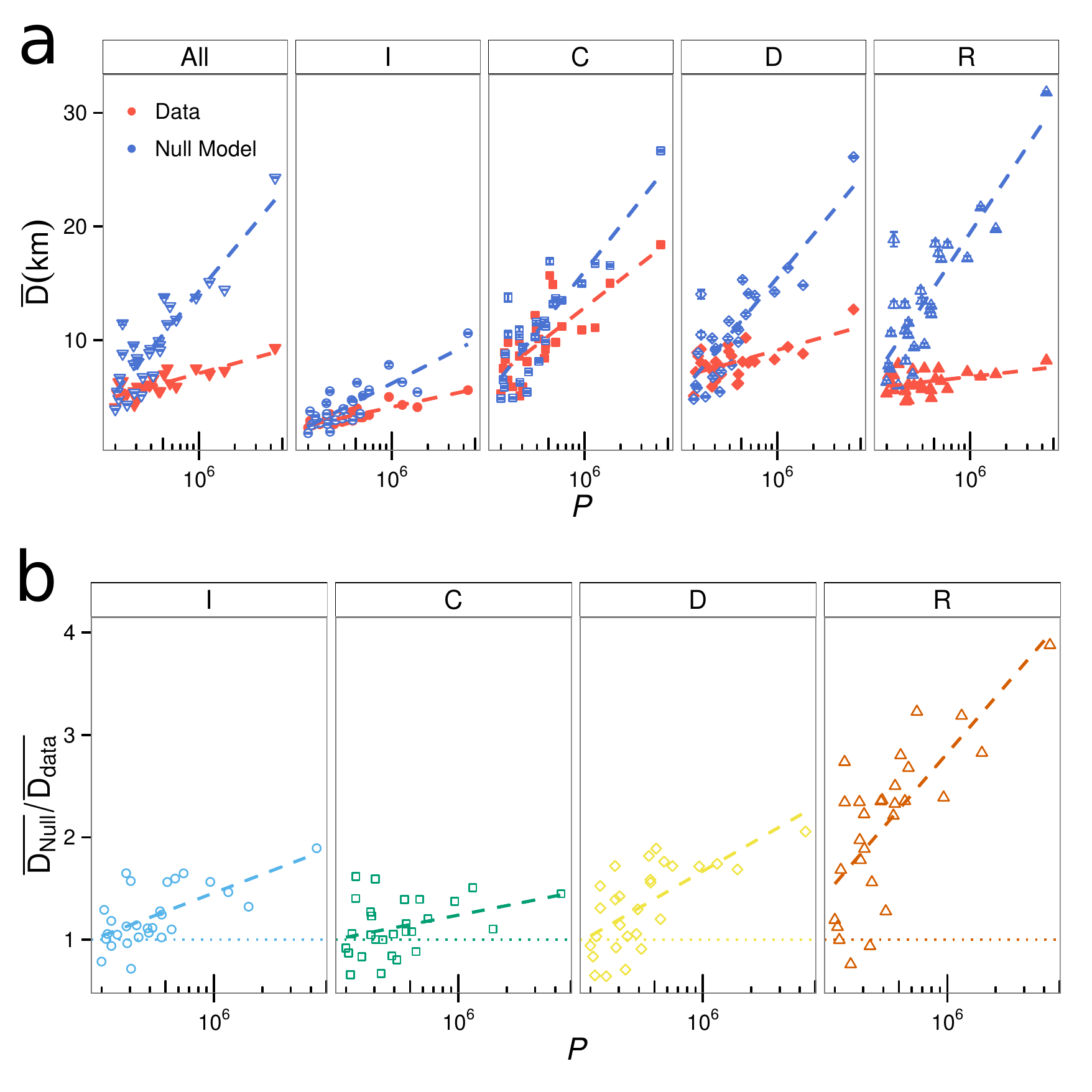}
  \caption{{\bf Distance per type of flow and population size.} (a)
    Average commuting distance vs. population size, per type of flows;
    (b) ratio $\overline{D_{Null}}/\overline{D_{Data}}$ per type of
    flows. The LouBar criteria~\cite{Louail:2014} is used here to
    define residential and employment hotspots.}
  \label{fig:avg-norm-commuting-dist}
\end{figure*}

We now want to characterize spatially these different flows and the
relation between city size and the commuting distances traveled by
individuals. In each city we compute the average distance traveled by
individuals per type of flows $I$,$C$,$D$ and $R$. The resulting
average distances measured in data are plotted in red on
Fig.~\ref{fig:avg-norm-commuting-dist}(a). We observe that the average
distance for all categories of flows increases with population size,
an expected effect as the city's area also grows with population
size. We also observe that the average distance of convergent flows
$C$ increases faster than for other types of flows (we note that the
average distance associated to convergent flows increases more than
the distance associated to divergent flows $D$, showing that these
flows are not symmetric as one could have naively expected). This
result means that for this set of Spanish cities, commuters from small
residential areas to important activity centers travel on average a
longer distance than all other individuals. This observation could be
an indication that for our set of cities, residential areas have
expanded while activity centers remained at their location, leading to
longer commuting distances (see Supplementary Figures
10 and 11 when considering various spatial scales of aggregation).

Another interesting information is provided by the comparison of
distances measured in the data with average distances measured from
the random OD matrices generated by the null model. The average
distances associated to the null model are plotted in blue in
Fig.~\ref{fig:avg-norm-commuting-dist}(a). We see that for all types
of flows the distances measured in the empirical data are shorter than
those generated by the null model. This is another clear indication of
the spatial organization of individual flows in cities. It also
highlights the importance of the travel time budget in the residential
locations choice. Remarkably enough, the distance of convergent flows
(C) is both the largest and the one that increases the fastest with
population, indicating a low degree of efficiency.

The comparison of this behavior with the null model leads to
interesting results. On Fig.~\ref{fig:avg-norm-commuting-dist}(b) we
plot the ratio $\overline{D_{Null}}/\overline{D_{Data}}$ for the four
types of flow. Values lower than $1$ indicate that the average
commuting distance generated by the null model is shorter than the
distance observed in the city.  Surprisingly, we observe that small
cities display a value less than one indicating the lesser importance
of space at this short scale. We also see that this ratio increases
faster for random flows ($R$) than for the others ($D,C,I$),
suggesting a remarkable spatial structure of these $R$ flows.

\begin{figure*}
  \centering
  \includegraphics[width=.8\linewidth]{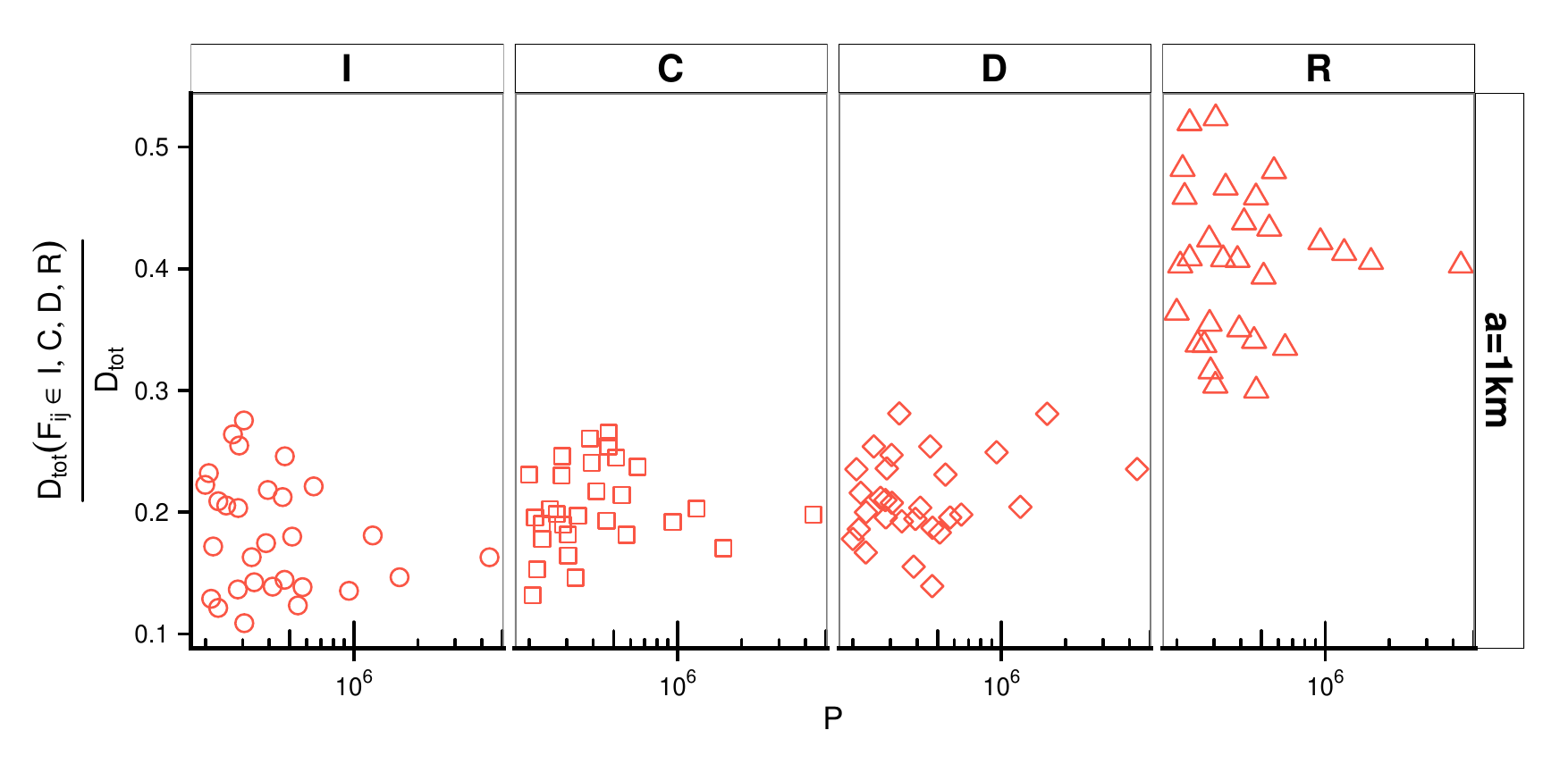}
  \caption{{\bf Total commuting distance.} Contributions to the total
    commuting distance of each type of flows vs. population size. We
    observe that the variations are small and that the largest
    contribution comes from the R flows.}
  \label{fig:total-commuting-dist}
\end{figure*}

We also consider the fraction of total commuting distance by type of
flow (Fig.~\ref{fig:total-commuting-dist}). We see that for each type
of flows, their respective fraction is constant and independent of
city size. With the LouBar hotspots detection method
\cite{Louail:2014} (see Supplementary Figure 3) and with a grid of
$1km^2$ square cells, we measure that roughly $40\%$ of the total
commuting distance is made on random flows while the other types
represent each about $20\%$. This result shows that the method is able
to identify where most of the commuting distance is traveled. In
particular, we see that the natural, obvious flows (I) from
residential centers to activity centers are not the most important
ones, and that the decentralization of commuting flows seems to be the
rule for the Spanish cities in our sample.

% Figure 4 : Contribution to total length of the 4 types of flow

\subsubsection*{Classification of cities}

\begin{figure}
  \centering
  \includegraphics{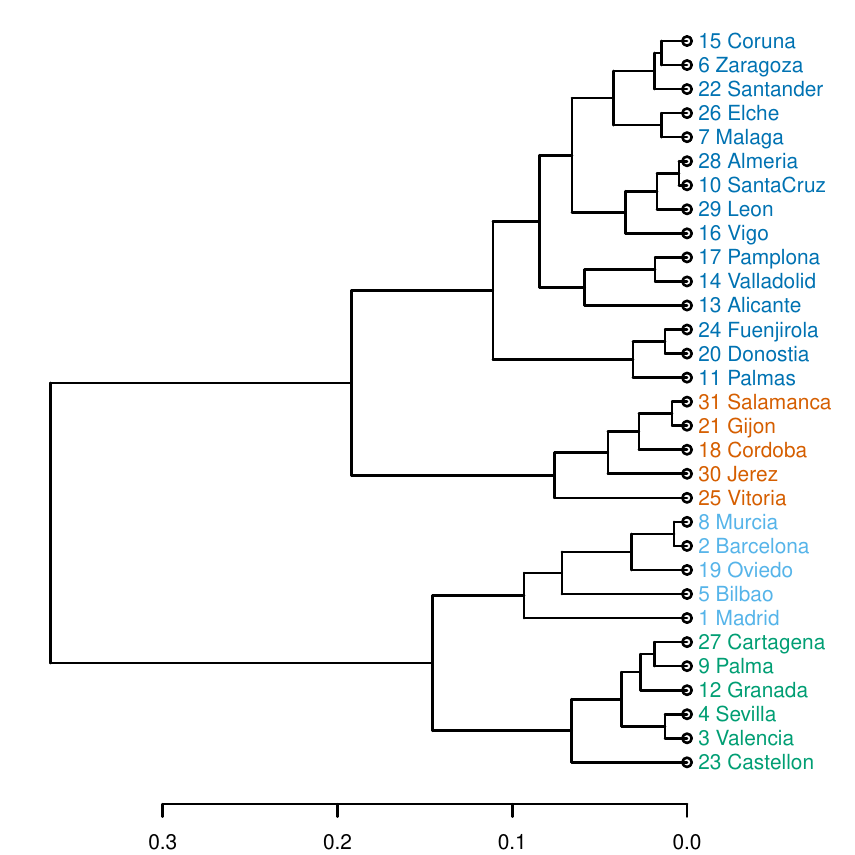}
  \caption{{\bf Classification of cities.} Dendrogram resulting from
    the hierarchical clustering on cities based on their ICDR
    values. In front of each city name we indicate its rank in the
    hierarchy of population sizes. The largest cities are clustered
    together. As cities get bigger, the `random' component ($R$) of
    their commuting flows increases, which signals that it is easier
    to commute from any place to any other in large cities.}
  \label{fig:dendrogram}
\end{figure}

Finally, the $ICDR$ signature of their OD matrix allows to cluster
cities with respect to the structure of their commuting patterns. We
measure the euclidian distance between the cities' $ICDR$ signatures
and we then perform a hierarchical cluster
analysis. Fig.~\ref{fig:dendrogram} shows the dendrogram resulting
from the classification. Four well-separated clusters are identified
on this dendrogram, and Table~\ref{table:classe-profiles} gives the
average value of each term along with the average population of the
cities composing the cluster. Remarkably these summary statistics show
that largest cities are clustered together and are characterized by a
larger proportion of `random' flows ($R$) of individuals both living
and working in parts of the city that are not the dominant residential
and activity centers. This can be interpreted as an increased facility
in bigger urban areas to commute from any part of the city to any
other part. Further studies on other cities and countries are needed
at this stage in order to discuss the relevance of the proposed
classification.

It is also important to test the robustness of this classification and
we show that introducing a reasonable amount of noise in the OD
matrices does not change the classification (see Methods and
Supplementary Figure 12). This sensitivity test confirms that the
clustering is robust against possible errors in the data source and in
the extraction of the mobility networks. The classification of cities
based on their $ICDR$ values is also reasonnably robust to a change of
the method used to define residential and employment hotspots (see
Supplementary Figure 13).

% Figure 5 : Dendrogram

\begin{table*}
  \centering
  \begin{tabular}{llccccc}
    Cluster & Cities & $\bar{P}$ & $\bar{I}$ & $\bar{R}$ & $\bar{D}$ & $\bar{C}$ \\
    \hline
    Orange & Salamanca, Gijon, Cordoba, \dots & $255,330$ & 0.43 & 0.27 & 0.16 & 0.14\\
    Dark blue & La Coru\~na, Zaragoza, Santander, Elche, \dots & $392,970$ & 0.37 & 0.36 & 0.15 & 0.13 \\
    Green & Cartagena, Palma, Granada, \dots & $732,992$ & 0.31 & 0.41 & 0.16 & 0.13 \\
    Light blue & Murcia, Barcelona, Bilbao, Madrid, \dots &
    $2,463,551$ & 0.25 & 0.46 & 0.17 & 0.12 \\
    \hline
  \end{tabular}
  \caption{{\bf Classification of cities.} Average $ICDR$ values and average population sizes of
    the cities composing each of the four clusters
    represented on Fig.~\ref{fig:dendrogram}. As the
    population grows the proportion of \textbf{R}andom flows increases while
    the proportion of \textbf{I}ntegrated decreases. The weights of
    \textbf{C}onvergent and \textbf{D}ivergent flows stay constant among the groups.}
  \label{table:classe-profiles}
\end{table*}

\section*{Discussion}

We have proposed a method to extract high-level information from large
weighted and directed networks, such as origin-destination
matrices. This method relies on the identification of origin and
destination hotspots, and this first step can be performed with any
reasonable method. The important second step consists in aggregating
flows in four different types, depending whether they start and end
from/to a hotspot or not.

We have applied this method to commuting networks extracted from
mobile phone data available in thirty-one Spanish cities. The method
has allowed us to highlight several remarkable patterns in the data:
\begin{itemize}

\item Independently of the density threshold chosen to determine
  hotspots, the proportion of integrated flows ($I$) decreases with
  city size, while the proportion of random flows ($R$) increases;

\item On average and for all cities considered here, individuals that
  live in residential main hubs and that work in employment main hubs
  ($I$ flows) travel shorter distances than the others ($C$,$D$,$R$
  flows); 

\item When the city size increases, the largest impact is on
  convergent flows ($C$) of individuals living in smaller residential
  areas (typically in the suburbs) and commuting to important
  employment centers;

\item The classification of cities based on the ICDR values leads to
  groups with consistent population size, highlighting a clear
  relationship between the population size of cities and their
  commuting structure.

\end{itemize}

In addition, the comparison with a null model led to interesting
conclusions. Flows in cities display a high level of spatial
organization and as the population size of the city grows, the
increase of the Z-scores of $I$,$C$,$D$,$R$ shows that the structure
of the mobility is more and more specific, and far from a random
organization. We note that an interesting direction for future
research would be to find some analytical arguments using simple
models of city organization for estimating these flows, and how they
vary with population size.

Our coarse-graining method provides a large scale picture of
individual flows in the city. In this respect it could be particularly
useful for validating synthetic results of urban mobility models (such
as \cite{Eubank:2004} for example), and for comparing different
models. An accurate modeling of mobility is indeed crucial in a large
number of applications, including the important case of epidemic
spreading which needs to be better understood, in particular at the
intra-urban level \cite{Balcan:2009,Dalziel:2013}.

It would also be interesting to apply the proposed method to other
mobility datasets, with different time and spatial scales, in order to
test the robustness of these commuting patterns. Another direction for
future studies could be to inspect the time evolution of the
$I$,$C$,$D$ and $R$ values for datasets describing travel to work
journeys over several decades (such as national travel surveys). The
method could also be applied at larger spatial scales, for example to
capture dominant effects in international migration flows. More
generally, we believe that an important feature of the ICDR method is
its versatility, as it could be applied to any type of data that is
naturally represented by a weighted and directed network.

\section*{Methods}

\subsection*{Data}
\label{sec:data}

The dataset used for our analysis comprises 55 days of aggregated and
anonymized records for 31 urban areas of more than 200,000
inhabitants. No individual information or records were available for
this study. The records included the set of Base Transceiver Stations
(communication antennas, BTS) used for the communications as presented
in the Call Detail Records (CDR). A CDR is produced for each active
phone event, including call/sms sending or reception. The number of
anonymized users represents on average $2\%$ of the total population
and at most $5\%$ of the total population. These percentages are
almost the same for all the urban areas. From the CDR data obtained
for $20$ weekdays (from mondays to thrusdays only), we extracted home
and work places for all the anonymized mobile phone users in the
dataset. The output of this processing phase is an OD
commuting matrix for each urban area, at the scale of the BTS point
pattern. In order to facilitate the calculations and the
comparison of the results between different cities, the OD matrices
are then transposed on regular square cells grids of varying size $a$
(see Supplementary Note 2).

\subsection*{Extraction of OD matrices from mobile phone data}
\label{sec:OD-extraction}

In order to extract OD matrices from phone calls, we select a subset
of users with a mobility displaying a sufficient level of statistical
regularity. For this analysis we considered commuting patterns during
workdays only. The users' Home and Work locations are identified as
the Voronoi cells which are the most frequently visited on weekdays by
each user between 8 pm and 7 am (Home) and between 9 am and 5 pm
(Work). We assume that there must be a daily travel between the home
and work locations of each individual. Users with a call activity
larger than $40\%$ of the days under study at home or work are
considered as valid. We then aggregate the complete flow of users and
construct the OD matrix with the flows between a Voronoi cell
classified as home and another cell classified as a work place. Since
the Voronoi areas do not exactly match the grid cells, we use a
transition matrix to change the spatial scale of the OD matrix, that
is to transform the $F_{ij}$ values of the OD matrix where $i$ and $j$
are Voronoi cells into $F'_{i'j'}$ values where $i'$ and $j'$ are the
cells of a regular grid (see Supplementary Note 2, and Supplementary
Figure 2 for an example of the partition an urban area in BTS Voronoi
cells).
    
\subsection*{Spatial scale of the OD matrix}
\label{sec:matrix-scale}

The OD matrix is the standard object in mobility studies and transport
planning~\cite{Ortuzar:1994} and contains information about movement
of individuals in a given area. More precisely, an OD matrix is a
$n*m$ matrix where $n$ is the number of different `Origin' zones, $m$
is the number of `Destination' zones and $F_{ij}$ is the number of
people commuting from place $i$ to place $j$ during a given period of
time. In transport surveys the size of the OD matrix depends on the
spatial scale at which the mobility data has been
collected. Traditionally the zones that are used to partition the city
are the administrative units, whose size can vary from census and
electoral units to whole departments or states, depending on the
purpose for building the OD matrix.

In this study we applied our ICDR method to cities divided in square
cells that are smaller than administrative units, allowing for a
better spatial resolution. In the case of OD matrices extracted from
CDR mobile phone data, the maximal resolution corresponds to the BTS
(antennas) point pattern. The $ICDR$ method proposed in this paper
does however not depend on a particular spatial scale and can be
applied on OD matrices available at coarser spatial resolutions as
well. For a given territory the results obtained with the ICDR method
- the I,C,D and R values in the first place - will obviously depend on
the spatial resolution and will also depend on the method used to
define hotspots. It is important to note that when the ICDR method is
used for comparing cities, the spatial resolution and the hotspots
identification method should be the same for all cities (see
Supplementary Notes 4 and 6, and Supplementary Figures 8 and 9 for
results obtained with another hotspots delimitation method, and
Supplementary Figure 7 for results obtained when considering another
spatial scale of aggregation).

\subsection*{Robustness of the classification of cities}

In order to ensure that the classification of cities based on their
ICDR matrices is robust, we introduce a noise in the flows $F_{ij}$
(for all the thirty-one cities). We focus on the case where the
workplace of an individual can be modified and where the number of
individuals living in each cell $i$ is kept constant. More precisely,
the noise is introduced as follows:
\begin{enumerate}

\item{} We pick up a uniform random positive integer $g$, the number of
  individuals whose workplace is reshuffled. This number $g$ varies
  from $1$ to $N=\sum F_{ij}$ the total number of commuters in the city.

\item{} We repeat $g$ times the following operation: we pick up randomly
  a residence and a workplace (a couple of values $(i,j)$) and move
  one individual from her workplace $j$ to put another randomly
  chosen workplace $j'$: $F_{ij} \rightarrow F_{ij}-1$ and $F_{ij'}
  \rightarrow F_{ij'}+1$

\end{enumerate}
The parameter of this workplace reshuffling is then $f=g/N$.  In order
to evaluate how much the classification of cities is affected by this
noise, we compute the Jaccard index $J_I$ between the reference
classification of cities in $k$ groups and the classification in $k$
groups obtained for the noisy OD matrices. The Jaccard index measures
the similarity of two partitions $P$ and $P'$ of same size:
\begin{equation}
J_I = \frac{a}{a+b+c}
\end{equation}
where
\begin{itemize}
\item $a$ : number of city pairs that are in the same group for both
  $P$ and $P'$
\item $b$ : number of city pairs that are in different groups in $P$
  but in the same one in $P'$
\item $c$ : number of city pairs that are in the same group in $P$ but
  not in $P'$ (or conversely)
\end{itemize}
The Jaccard index $J_I$ is in the range $[0;1]$ and the closest it is
to 1, the larger the similarity between $P$ and $P'$.

We generate $100$ noisy matrices for each value of $f$ and compute the
average value $\bar{J_I}$ of the Jaccard index. This average value
encodes the distance between the reference partition $P$ of cities in
$k$ groups and the partitions of cities in $k$ groups obtained for the
noisy OD matrices. Supplementary Figure 12 shows the values of
$\bar{J_I}$ versus the proportion $f$ of reshuffled individuals, for
different number of groups $k$. The red shaded rectangle on each panel
corresponds to the mean value +/- the standard deviation obtained for
$1,000$ replications of a null model, in which permutations of cities
among the $k$ groups are randomly performed. We observe here that up
to $20\%$ of reshuffled individuals, the average value $\bar{J_I}$
obtained is always significantly larger than the one obtained for the
null model, indicating that the classification is robust even for
important values of the noise.

%% Optional Appendix or Appendices
%% \appendix Appendix text...
%% or, for appendix with title, use square brackets:
%% \appendix[Appendix Title]

\section*{Acknowledgments}
We thank the anonymous referees for interesting and constructive
comments. The research leading to these results has received funding
from the European Union Seventh Framework Programme FP7/2007-2013
under grant agreement 318367 (EUNOIA project).  ML and JJR received
partial financial support from the Spanish Ministry of Economy
(MINECO) and FEDER (EU) under projects MODASS (FIS2011-24785) and
INTENSE@COSYP (FIS2012-30634). The work of ML has been funded under
the PD/004/2013 project, from the Conselleria de Educación, Cultura y
Universidades of the Government of the Balearic Islands and from the
European Social Fund through the Balearic Islands ESF operational
program for 2013-2017.

\section*{Author contributions}

T.L. designed the study, processed and analyzed the data and wrote the
manuscript; M.L. processed and analyzed the data; O.G.C and
M.P. processed the data; R.H. and J.J.R. coordinated the study;
E.F.-M. obtained and processed the data; M.B. coordinated and designed
the study, and wrote the manuscript. All authors read, commented and
approved the final version of the manuscript.

\section*{Additional information}

\textbf{Supplementary Information} accompanies this paper.

\smallskip 

\textbf{Competing financial interests:}~The authors declare no
competing financial interests.

%%%%%%%%%%%%%%%%%%%%%%%%%%%%%%%%%%%%%%%%%%%%%%%%%%%%%%%%%%%%%%%%

%\pagebreak

\clearpage

\section*{Supplementary Information}

%\maketitle
\subsection*{Supplementary Figures}

\begin{figure}[h]
 \centering
  \includegraphics{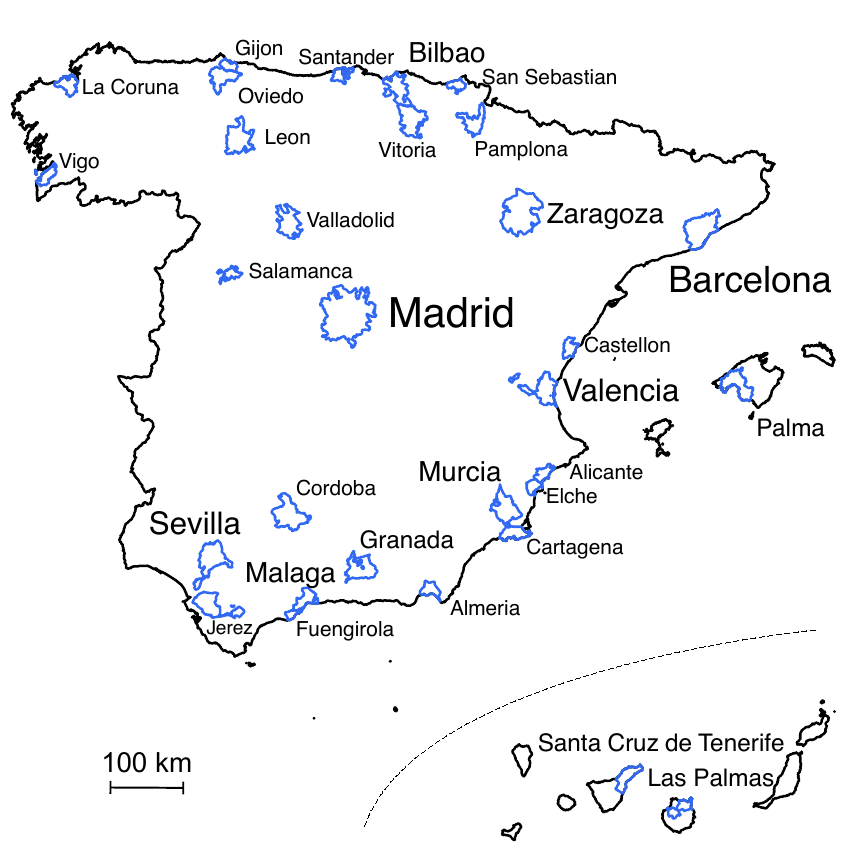}
\captionsetup{labelformat=empty}{Figure S1. {\bf The 31 biggest Spanish urban areas under study.}}
  \label{fig:map-Spain}
\end{figure}

%\clearpage

\begin{figure}
  %\centering
  \includegraphics[width=0.9\linewidth]{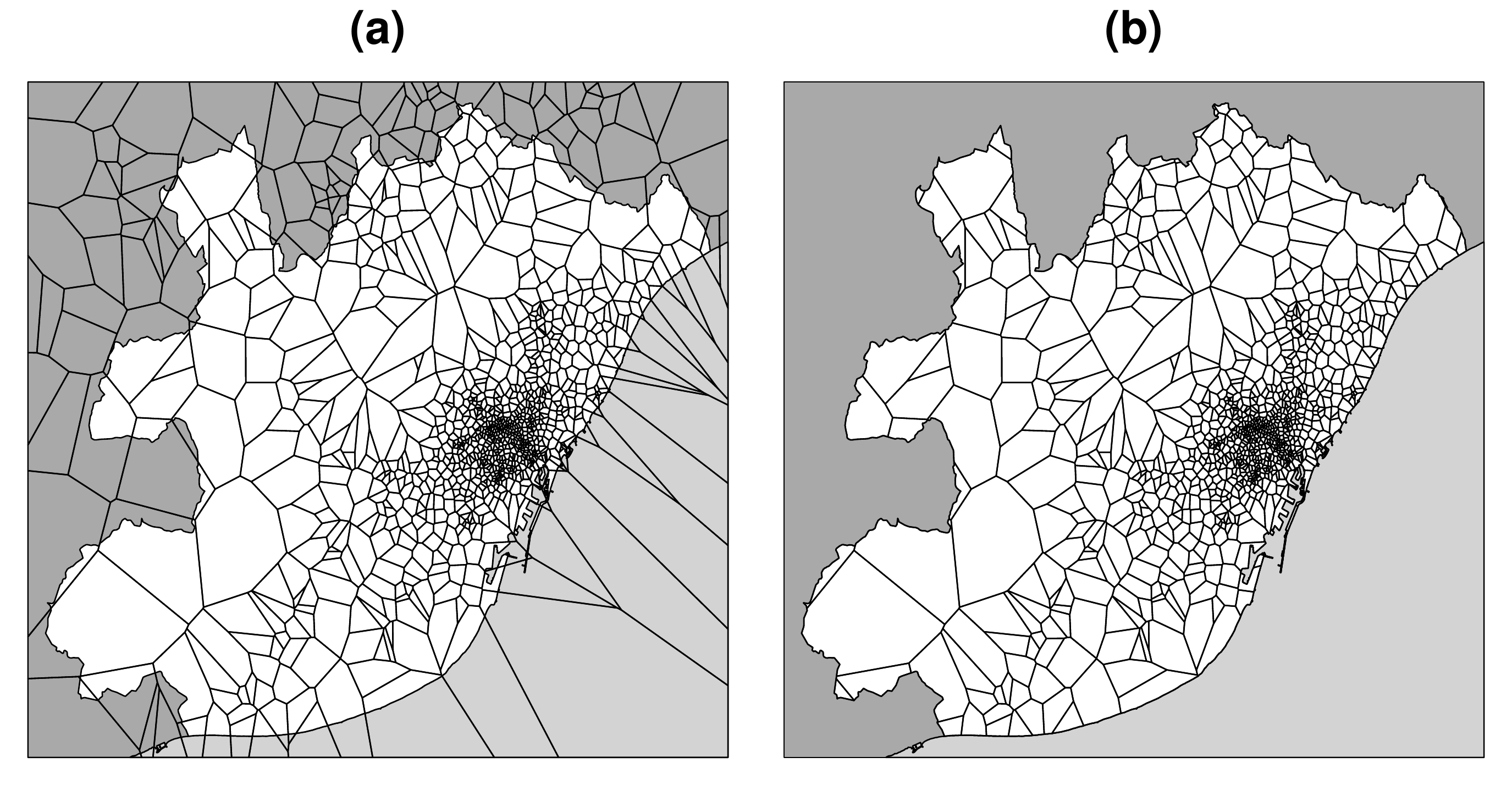}
 \captionsetup{labelformat=empty}{Figure S2. {\bf Map of the metropolitan area of
    Barcelona.} The white area represents the
    metropolitan area, the dark grey area represents
    territory surrounding the metropolitan area and the
    gray area the sea. (a) Voronoi cells. (b)
    Intersection between the Voronoi cells and the
    metropolitan area.\label{fig:Voronoi}}
\end{figure}

%\clearpage

\begin{figure}
%  \centering
  \includegraphics[width=0.9\linewidth]{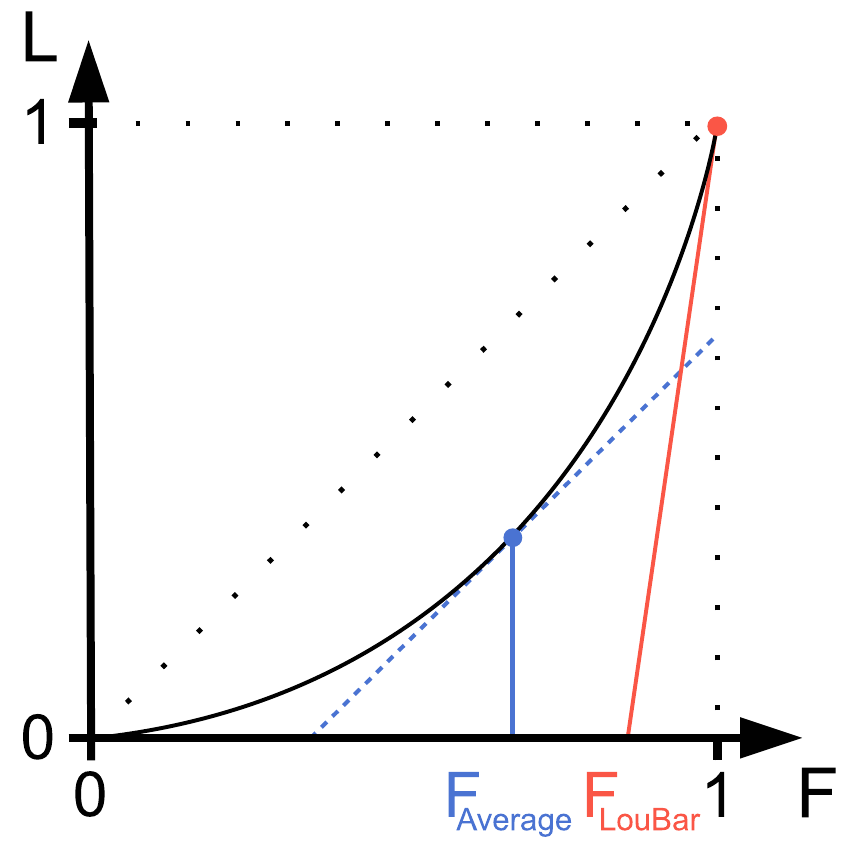}
\captionsetup{labelformat=empty}{Figure S3. {\bf Illustration of the `LouBar' hotspots determination
      method.} We show the Lorenz curve and how we determine the
    thresholds for selecting hotspots (see details in Supplementary
    Note \ref{sec:hotspots-def}).}
  \label{fig:Lorenz-curve}
\end{figure}

%\clearpage

\begin{figure}
 \centering
 \begin{tabular}{cc}
   (a) & (b) \\
   \includegraphics[width=0.45\linewidth]{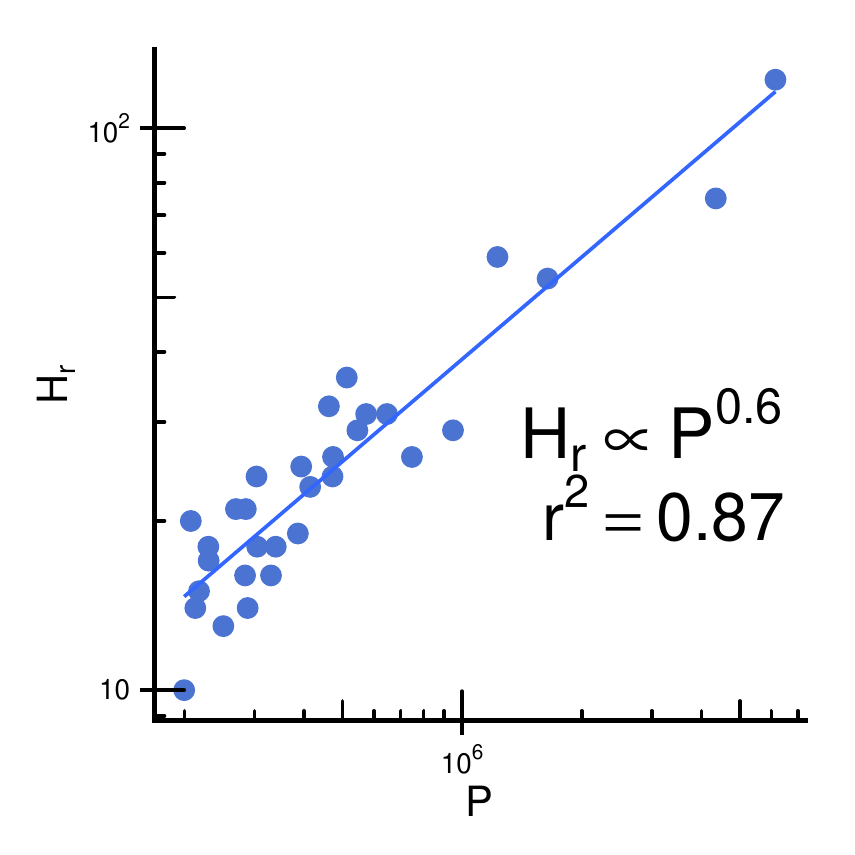} & 
   \includegraphics[width=0.45\linewidth]{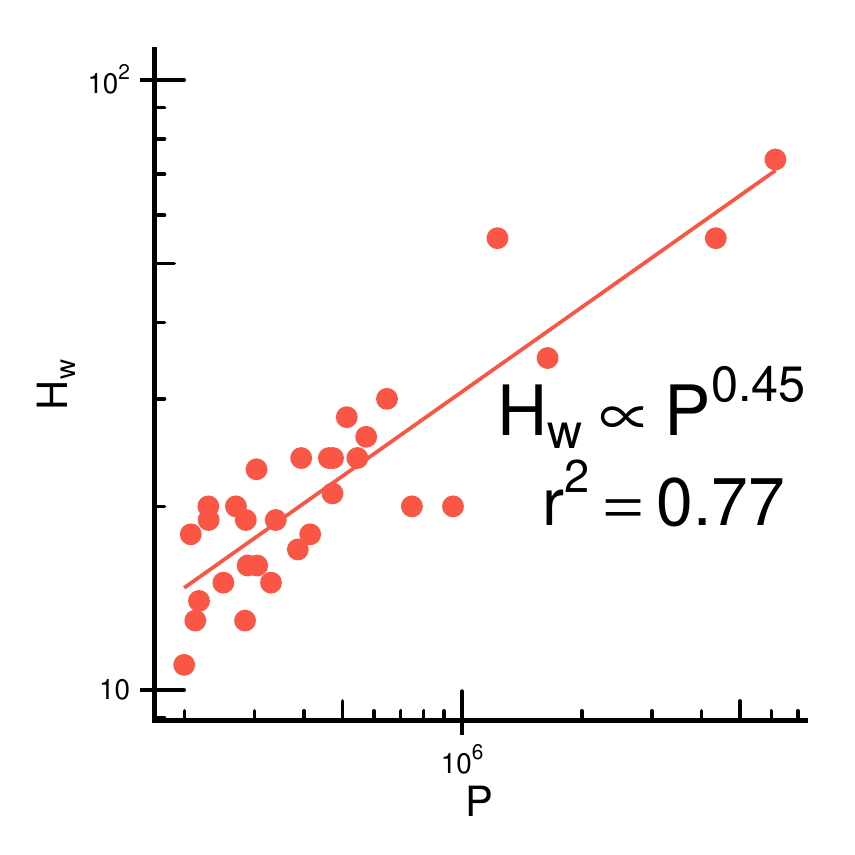} 
 \end{tabular}
 \captionsetup{labelformat=empty}{Figure S4. {\bf Scaling of hotspots.} (a) Number of residential
   hotspots vs. population size of the city ; (b) Number of work
   hotspots vs. population size of the city ; The scaling relation is
   sublinear, indicating an 'economy of scale' in the spatial
   organisation of cities. The exponent is remarkably smaller in the
   case of the work/main daily activity hotspots, which means that in
   Spanish cities the number of important places that concentrate many
   jobs and daily amenities, grows slower than the number of important
   residential places.}
 \label{fig:n-hotspots}
\end{figure}

%\clearpage

\begin{figure}
 \centering
 \includegraphics[width=\linewidth]{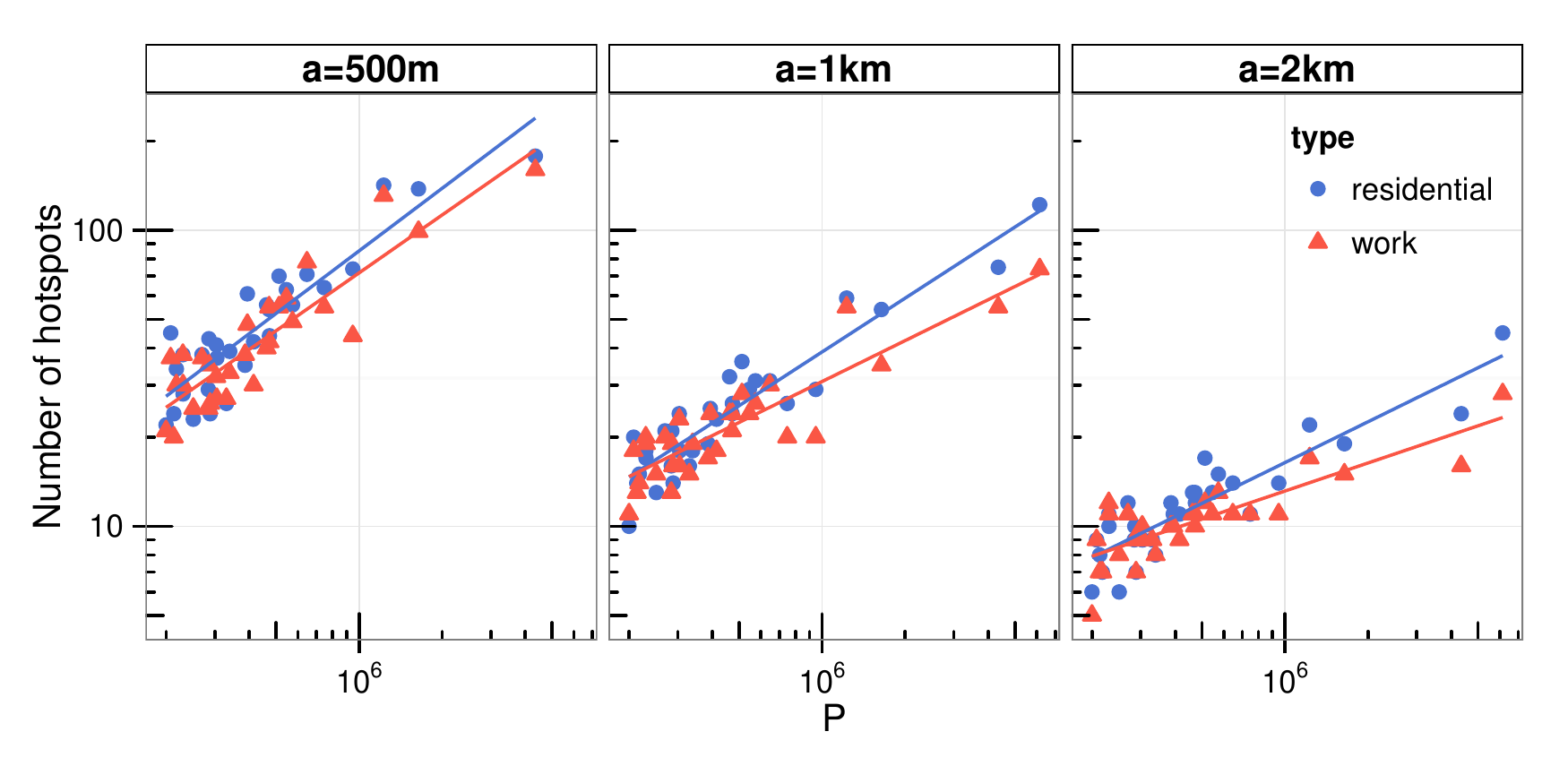}
 \captionsetup{labelformat=empty}{Figure S5. {\bf Effect of grid size on
     scalings.} Numbers of residential hotspots and work hotspots
   (defined with the LouBar method, see
   section~\ref{sec:hotspots-def}) vs. the population size of the city
   for different sizes $a$ of grid cells.  In any case the scaling
   relations for both quantities stay sublinear, indicating an
   "economy of scale" in the spatial organisation of cities. Also
   remarkable is that the scaling exponent of the number of
   residential centers is always larger than the exponent of work
   centers.}
\label{fig:n-hotspots-var-a}
\end{figure}

%\clearpage

\begin{figure}
  \centering
  \begin{tabular}{cc}
    (a) & (b) \\
    \includegraphics[width=0.45\linewidth]{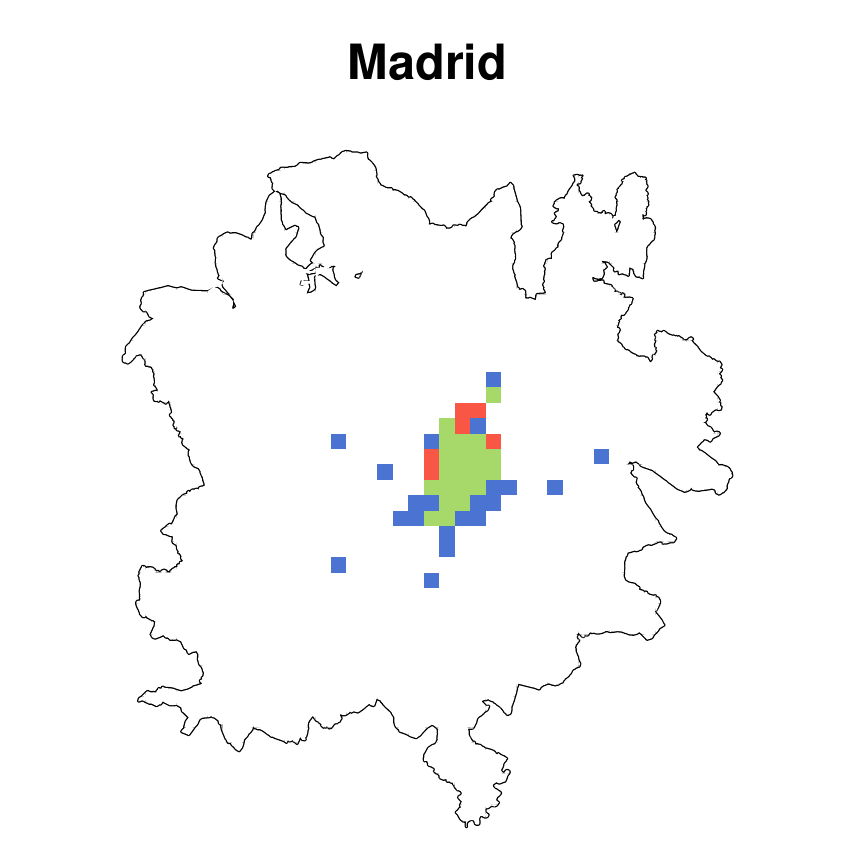} &
    \includegraphics[width=0.45\linewidth]{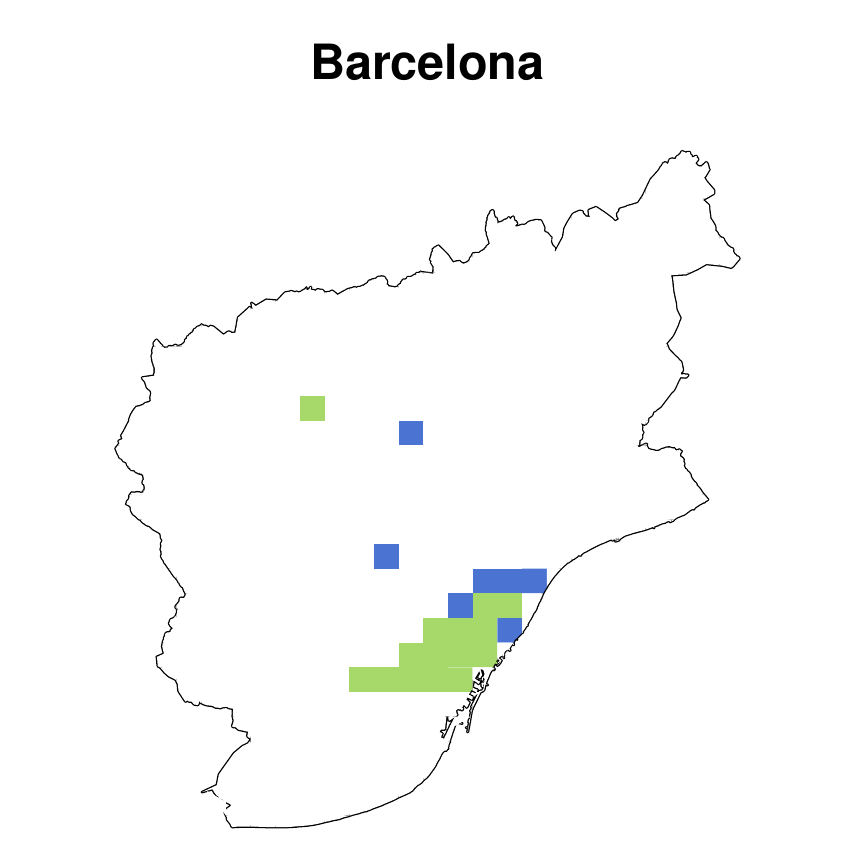} \\
    (c) & (d) \\
    \includegraphics[width=0.45\linewidth]{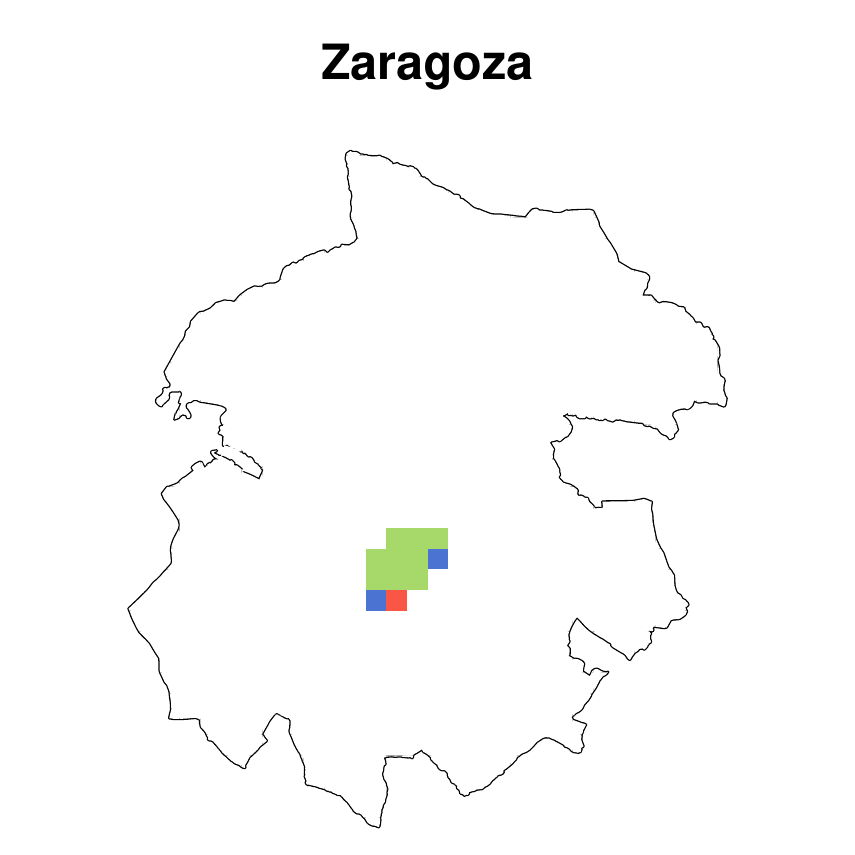} &
    \includegraphics[width=0.45\linewidth]{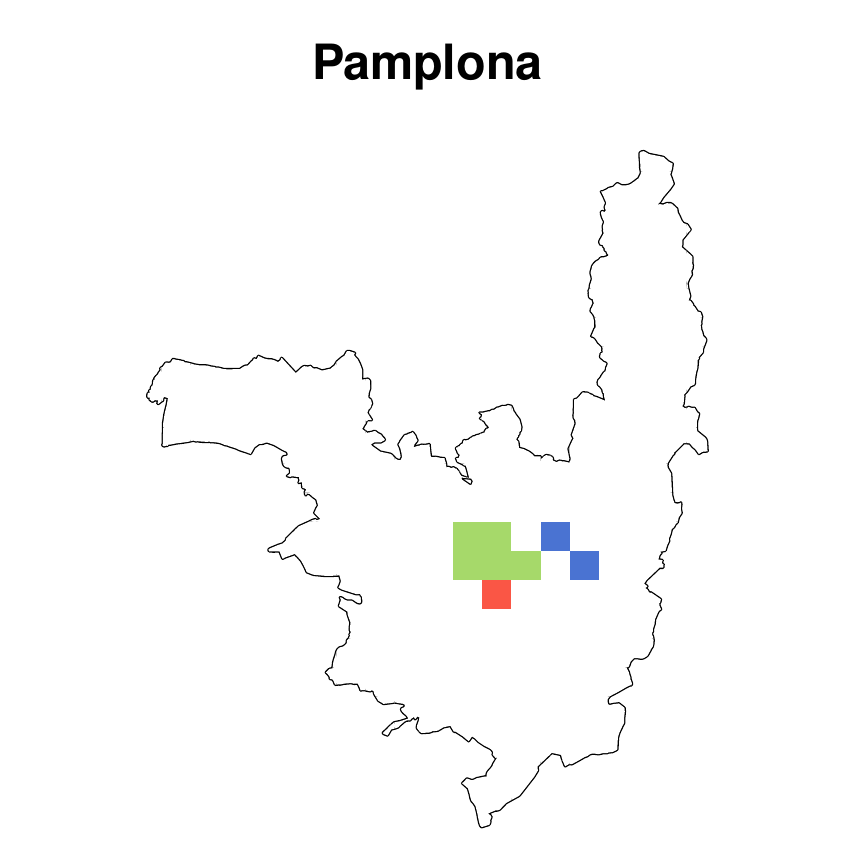} \\
  \end{tabular}
  \captionsetup{labelformat=empty}{Figure S6. {\bf Location maps of residential
      and work hotspots in four cities.} Each square cell represents a
    zone of $4~km^2$. Blue cells are residential hotspots only, red
    cells are work hotspots only, and green cells are simultaneously
    residential and work hotspots. These maps exhibit different
    spatial organisations of residential and work hotspots.}
  \label{fig:hotspots-maps}
\end{figure}

%\clearpage

\begin{figure}
  \centering
  \includegraphics{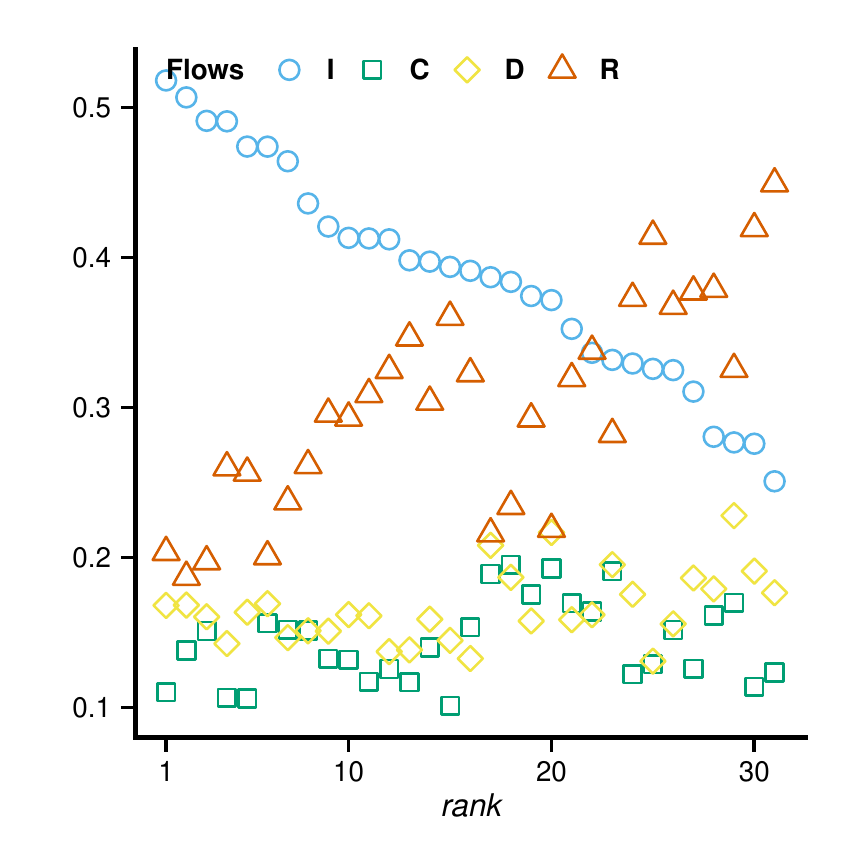}
 \captionsetup{labelformat=empty}{Figure S7. {\bf Effect of grid size on ICDR values.} $I$
    (integrated),$C$ (convergent),$D$ (divergent) and $R$ (random)
    values of the 31 Spanish urban areas ranked by decreasing values
    of $I$, for $a=2km$. While more dispersed, the data exhibit the
    same pattern than the one obtained with $a=1km$ and represented on
    Figure 2 in the main text.}
  \label{fig:ICDR-LouBar-2km}
\end{figure}

%\clearpage

\begin{figure}
  \centering
  \includegraphics[width=\linewidth]{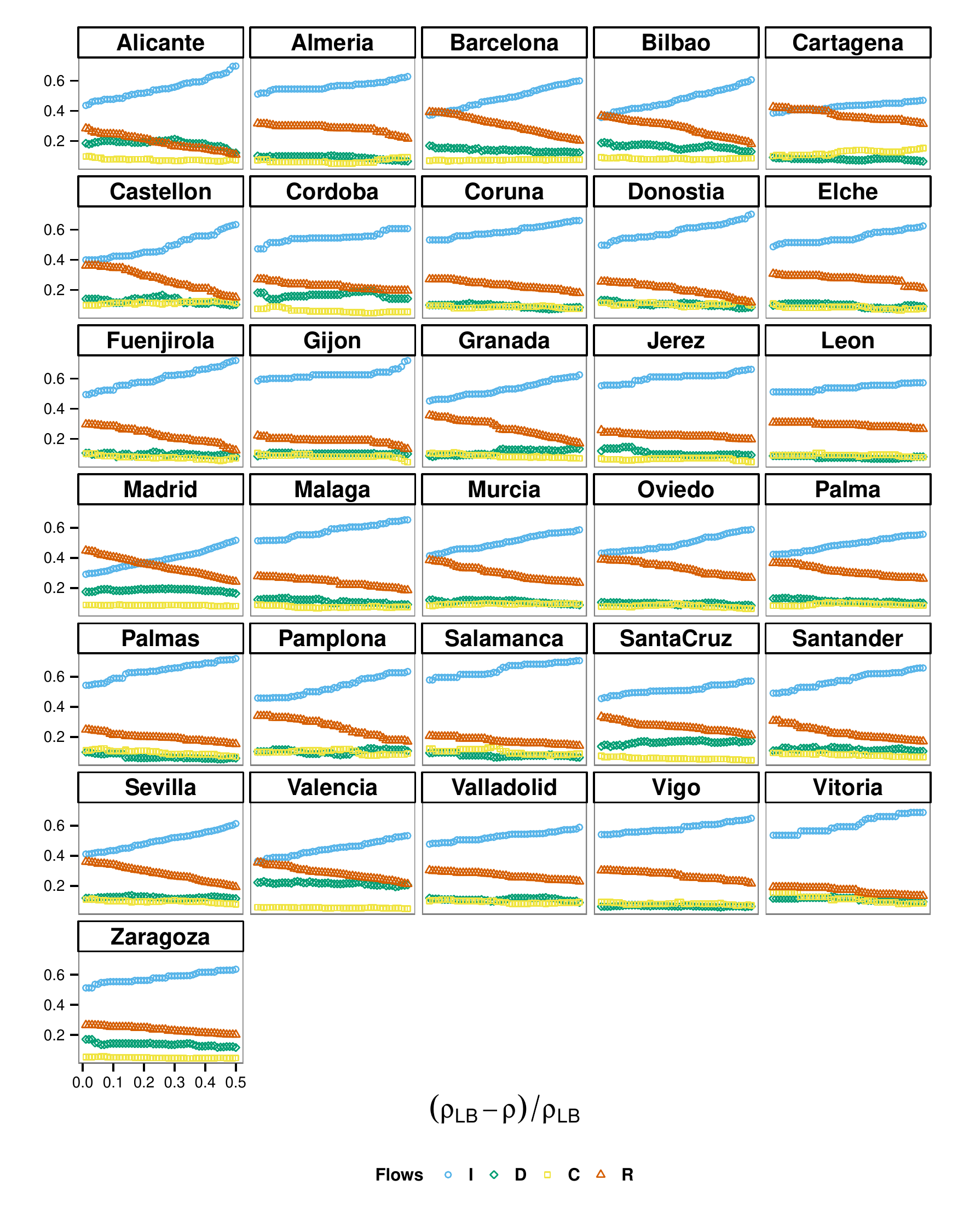}
 \captionsetup{labelformat=empty}{Figure S8. {\bf Sensitivity of ICDR values
      with the density threshold.} $I$ (integrated),$C$
    (convergent),$D$ (divergent) and $R$ (random) values of the 31
    Spanish cities ranked by decreasing as a function of the density
    threshold chosen to determine origin and destination hotspots of
    the network. }
  \label{fig:ICDR-sensitivity-1km}
\end{figure}

%\clearpage

\begin{figure}
  \centering
  \includegraphics{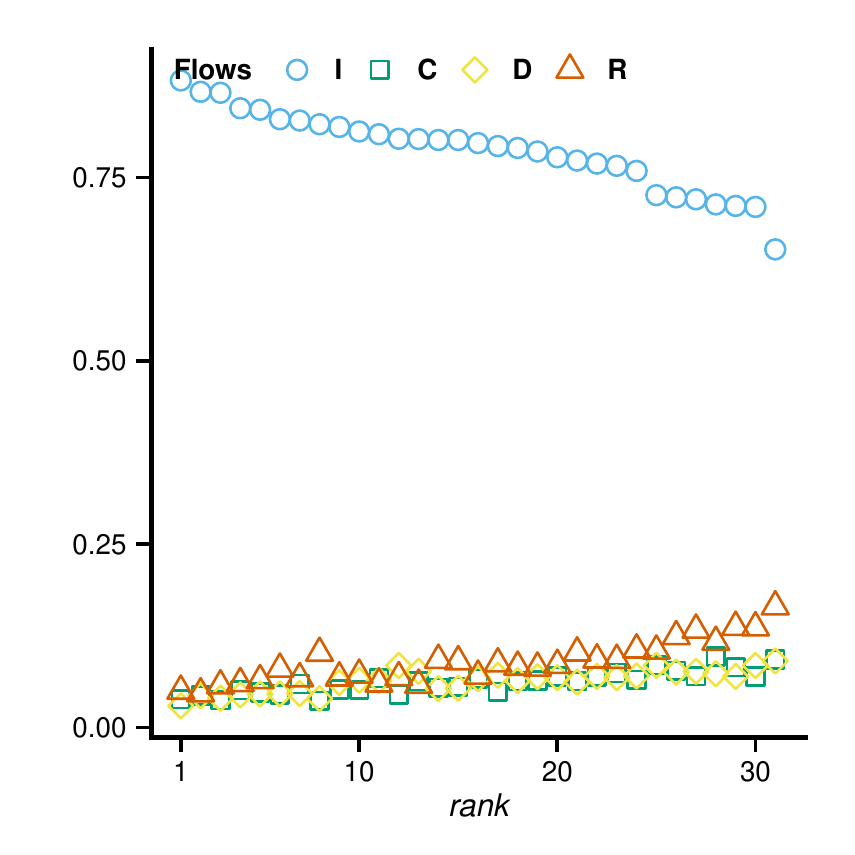}
  \captionsetup{labelformat=empty}{Figure S9. {\bf ICDR values obtained with the 'Average' criteria.} $I$
    (integrated),$C$ (convergent),$D$ (divergent) and $R$ (random)
    values of the 31 Spanish cities ranked by decreasing order of $I$,
    for $a=1km$ and when defining hotspots with respect to the
    'Average' criteria (i.e. all cells whose density of
    residents/workers is greater than the mean value are considered as
    residential/employment hotspots.}
  \label{fig:ICDR-Average-1km}
\end{figure}

%\clearpage

\begin{figure}
  \centering
  \includegraphics[width=\linewidth]{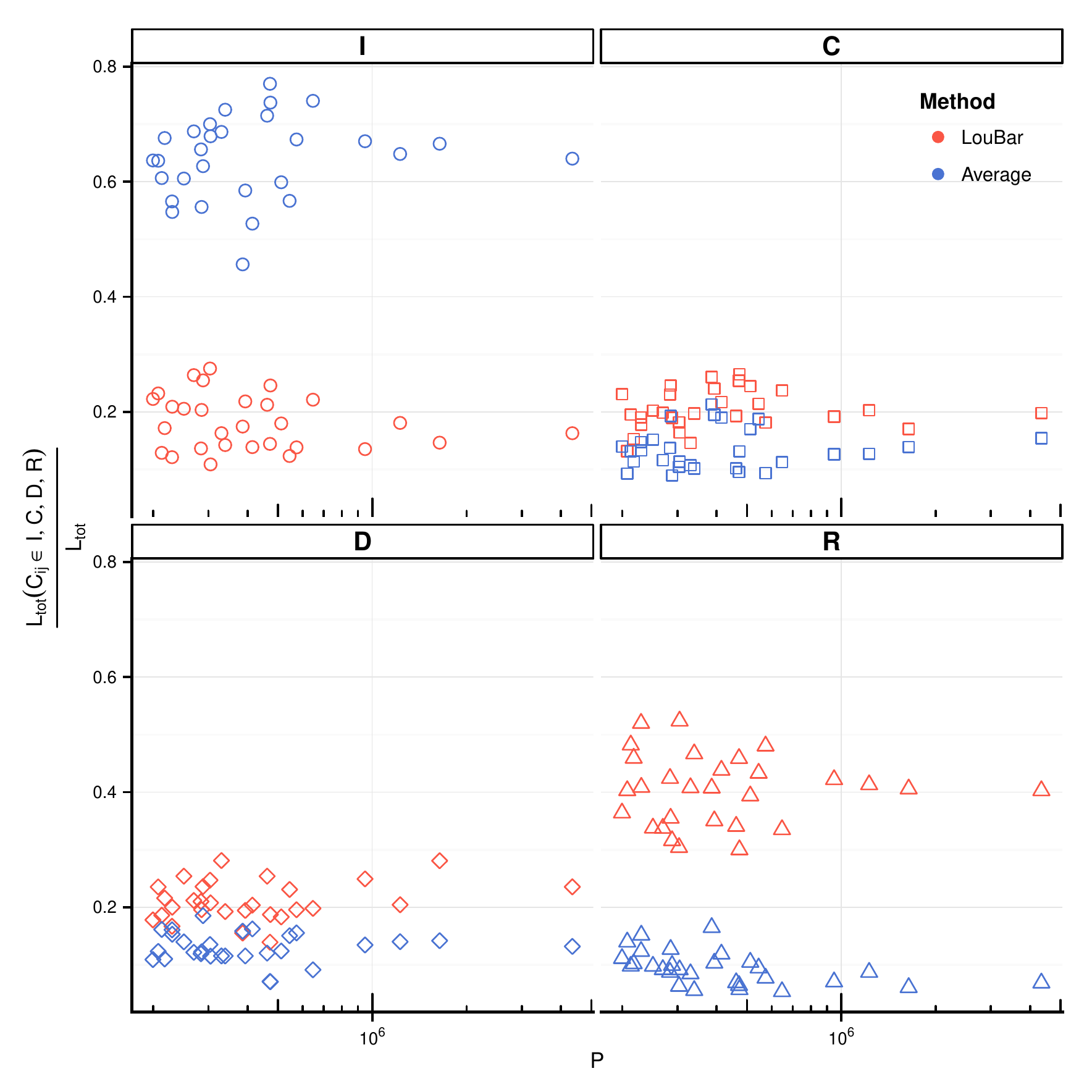}
  \captionsetup{labelformat=empty}{Figure S10.  {\bf Commuting distance per flow type ($a=1$km).}
    Contribution to the total commuting distance of each type of flow
    vs. population size, for two different hotspots definition
    methods, and for $a=1km$. These contributions do not depend on
    population size.}
  \label{fig:total-norm-dist-1000}
\end{figure}

%\clearpage

\begin{figure}
  \centering
  \includegraphics[width=\linewidth]{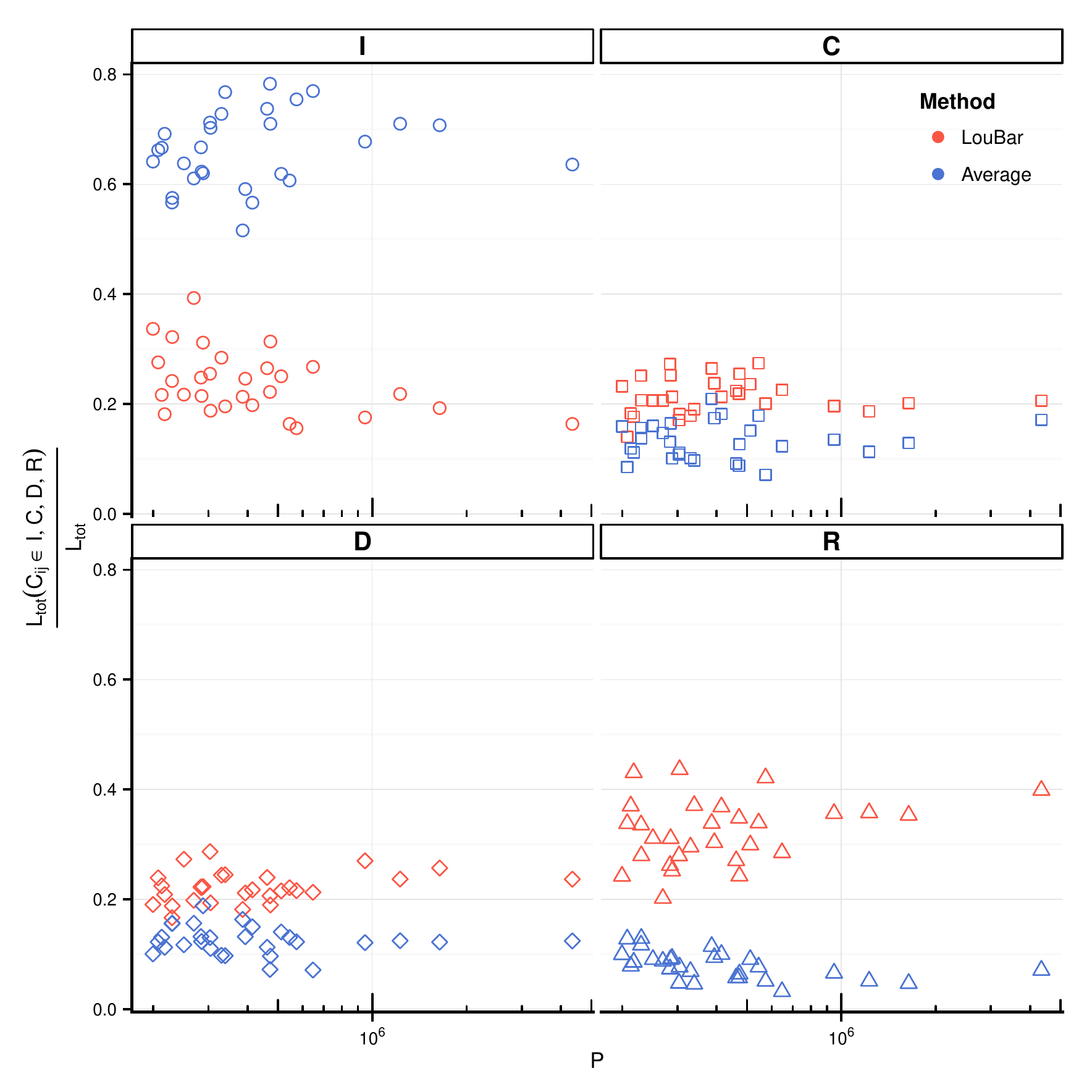}
  \captionsetup{labelformat=empty}{Figure S11. {\bf Commuting distance per flow type ($a=2$km).}
    Contribution to the total commuting distance of each type of flow
    vs. population size, for two different hotspots definition
    methods, and for $a=2km$. These contributions do not depend on
    population.}
  \label{fig:total-norm-dist-2000}
\end{figure}

%\clearpage

\begin{figure}
  \centering
  \includegraphics[width=\linewidth]{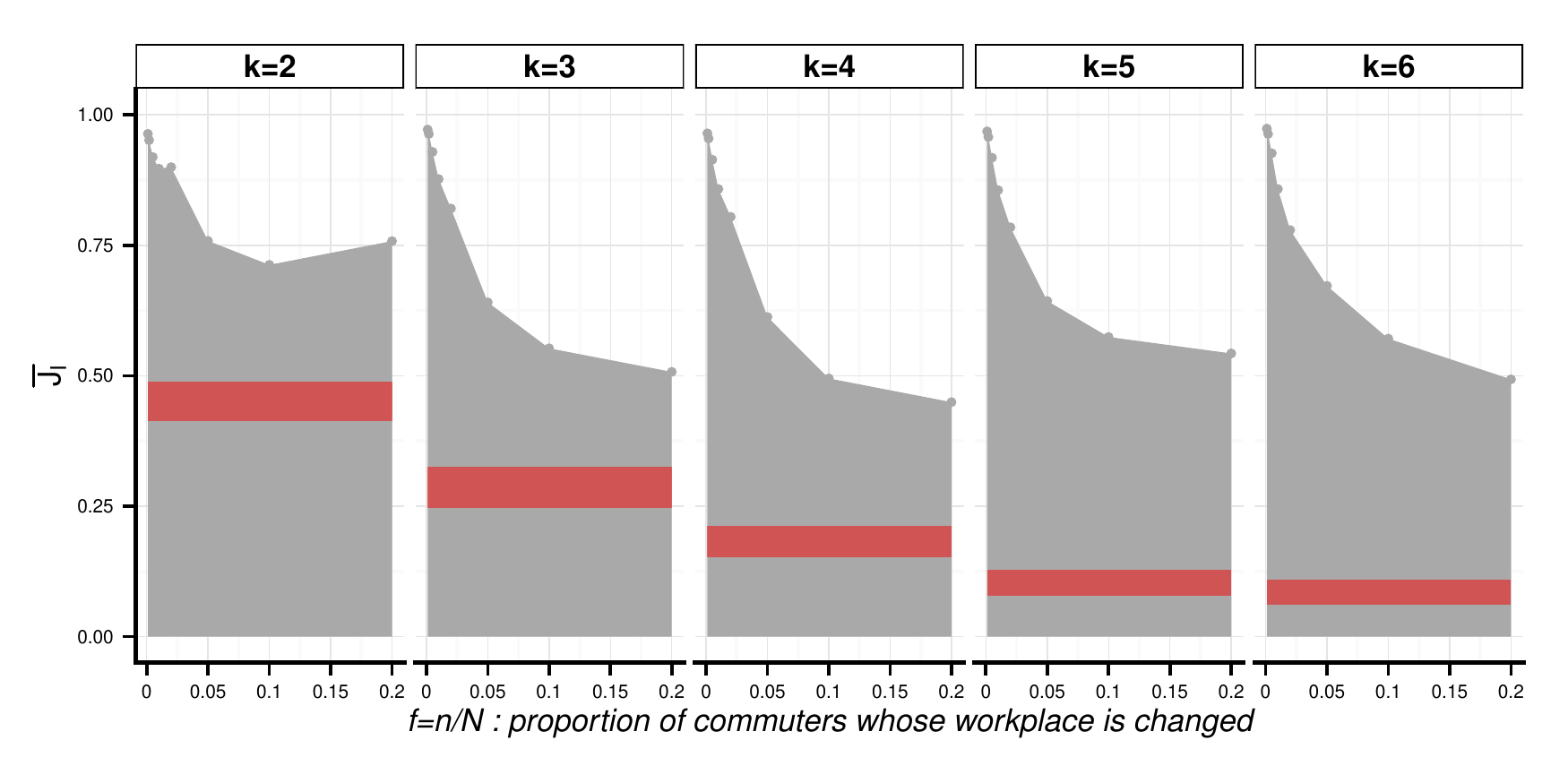}
  \captionsetup{labelformat=empty}{Figure S12. {\bf Effect of noise on clusters.} Jaccard index $J_I$
    resuming the sensitivity of the clusters to some uniform noise
    introduced in the commuting flows for different numbers of groups
    $k$. The light red shaded rectangle on each panel corresponds to
    the mean value +/- the standard deviation obtained for 1000
    replications of a null model, in which permutations of cities
    among the $k$ groups are performed randomly.}
  \label{fig:jaccard-ICDR-noise}
\end{figure}

%\clearpage

\begin{figure}
  \centering
  \includegraphics[width=\linewidth]{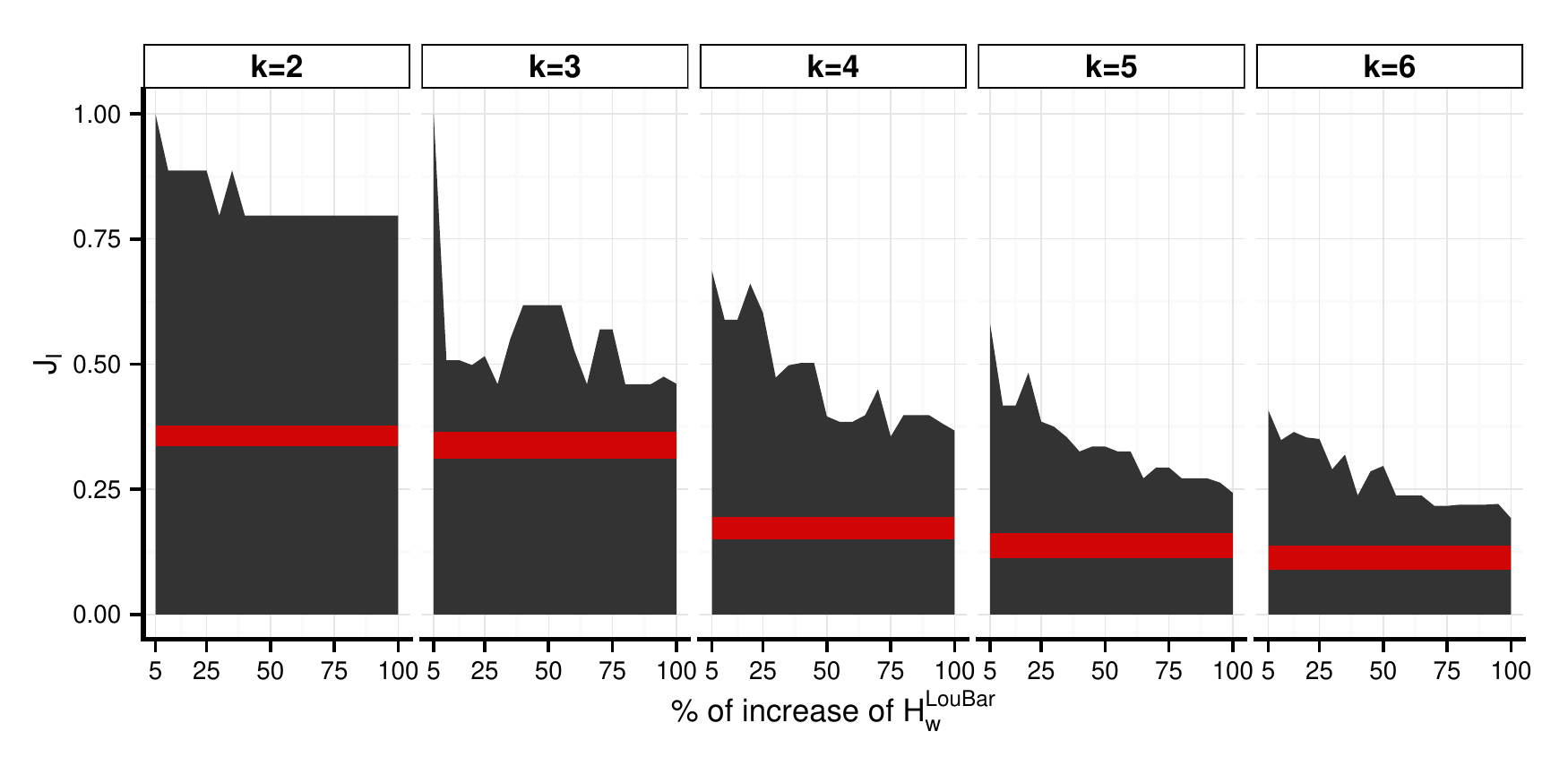}
 \captionsetup{labelformat=empty}{Figure S13. {\bf Effect of hotspots definition on clusters.} Jaccard
    index $J_I$ resuming the sensitivity of the clusters to the
    threshold chosen to determine hotspots. The light red shaded
    rectangle on each panel corresponds to the mean value +/- the
    standard deviation obtained for 1000 replications of a null model,
    in which permutations of cities among the $k$ groups are performed
    randomly.}
  \label{fig:jaccard-hotspots-def}
\end{figure}

\clearpage

%\section{Supplementary Tables}

\subsection*{Supplementary Notes}

\subsubsection*{Supplementary note 1}
%. Selection of cities and definition of the boundaries of cities}
\label{sec:city-def}

Comparing properties of cities of very different population sizes and
areas requires to rely on a harmonized, spatial delimitation of cities
that goes beyond the administrative boundaries defined by national
authorities. These administrative boundaries may be somewhat arbitrary
and do not follow neither the morphological nor functionnal spatial
footprint of the city. To this end we have chosen to rely on the
\emph{urban areas} defined by the AUDES initiative (Areas Urbanas De
ESpa\~na: AUDES project. Documentation and open data
  available at \url{http://alarcos.esi.uclm.es/per/fruiz/audes/} (accessed
  January 27, 2014)). This project is an attempt to capture some
coherent spatial delimitations of Spanish cities regarding the
(journey to work) commuting patterns of individuals living in the core
city of each urban area and in their surrounding municipalities. These
delimitations are built upon statistical criteria based on the
proportion of residents of surrounding municipalities that commute to
the main city to work. The locations and spatial boundaries of the 31
Spanish urban areas analysed in this study are represented on
Supplementary Figure S1.%~\ref{fig:map-Spain}.

\subsubsection*{Supplementary note 2}
%Mobile phone data pre-processing}
\label{sec:data-processing}

\paragraph*{Voronoi cells} We remove the BTSs with zero mobile phone
users and we compute the Voronoi cells associated with each BTSs of
the urban area (hereafter called UA). We remark in Supplementary
Figure S2a that there are four types of Voronoi cells:

\begin{enumerate}
\item The Voronoi cells contained in the UA
\item The Voronoi cells partitionned between the UA and the
  territory outside the UA.
\item The Voronoi cells between the UA and the sea (noted S).
\item The Voronoi cells between the UA, the territory outside the
  UA and the sea.
\end{enumerate}

To compute the number of users associated with the intersections
between the Voronoi cells and the UA we have to take into account
these different types of Voronoi cells. Let $m$ be the number of
Voronoi cells, $N_{v}$ the number of mobile phone users in the Voronoi
cell $v$ and $A_{v}$ the area of the Voronoi cell $v$, $v \in
|[1,m]|$. The number of users $N_{v\cap UA}$ in the intersection
between $v$ and UA is given by the following equation:

\begin{equation}
  N_{v\cap UA}=N_v \left(\frac{\displaystyle A_{v\cap UA}}{A_v - A_{v\cap S}}\right)
  \label{eq:vUA}
\end{equation}

We note in Equation \ref{eq:vUA} that we remove the intersection of
the Voronoi area with the sea, indeed, we assume that the number of
users calling from the sea are negligeable. Now we consider the number
of mobile phone users $N_v$ and the associated area $A_v$ of the
Voronoi cells intersecting the UA (see Supplementary Figure S2b).

\paragraph*{Extraction of Origin-Destination matrices}

%\subsubsection{From BTS point patterns to regular grids of square cells}
We cannot directly extract an OD matrix between the grid cells with
the mobile phone data because each users' home and work locations are
identified by the Voronoi cells. Thus, we need a transition matrix $P$
to transform the BTS OD matrix $B$ into a grid OD matrix $G$.

Let $m$ be the number of Voronoi cells covering the urban area and $n$
be the number of grid cells. Let $B$ be the OD matrix between BTSs
where $B_{ij}$ is the number of commuters between the BTS $i$ and the
BTS $j$. To transform the matrix $B$ into an OD matrix between grid
cells $G$ we define the transition matrix $P$ where $P_{ij}$ is the
area of the intersection between the grid cell $i$ and the BTS
$j$. Then we normalize $P$ by column in order to consider a proportion
of the BTSs areas instead of an absolut value, thus we obtain a new
matrix $\hat{P}$ (Equation \ref{eq:pchap}).

\begin{equation}
	\hat{P}_{ij}=\frac{\displaystyle P_{ij}}{\sum_{k=1}^m \displaystyle P_{kj}}
	\label{eq:pchap}
\end{equation}

The OD matrix between the grid cells $G$ is obtained by a matrices
multiplication given in the following equation:

\begin{equation}
	 G=P B P^t
	\label{OD}
\end{equation}

\subsubsection*{Supplementary note 3}
%Definition of residential and employment hotspots}
\label{sec:hotspots-def}

A city is divided in $n$ cells and the data give us access to its
$n\times n$ OD matrix extracted from the mobile phone data. After
straightforward calculation we obtain the distributions $(C_R)$ and
$(C_W)$ of the numbers of residents/workers in the $n$ cells composing
the city.  The determination of centres and subcentres is a problem
which has been broadly tackled in urban economics (see for example the
references given in the paper and references in \cite{Louail:2014}).
Starting from a spatial distribution of densities, the goal is to
identify the local maxima. This is in principle a simple problem
solved by the choice of a threshold $\delta$ for the density $\rho$: a
cell $i$ is a hotspot if the density of users $\rho(i) >\delta$. It is
however clear that such method based on a fixed threshold introduce
some arbitrariness due to the choice of $\delta$, and also requires
prior knowledge of the city to which it is applied to choose a
relevant value of $\delta$. In \cite{Louail:2014} we proposed a
generic method to determine hotspots from the Lorenz curve of the
densities. In the following we quickly introduce the principle of the
method and its application to the determination of the residential and
work hotspots of each city. We invite the interested readers to refer
to \cite{Louail:2014} for further discussion on this method.

\smallskip

We first sort $(C_R)$ and $(C_W)$ in increasing order, and denote the
ranked values by $C_R^1 (resp.~C_W^1) <C_R^2 < ... <C_R^n$ where $n$
is the number of cells.  The two Lorenz curves of the distribution of
residents/workers are constructed by plotting on the x-axis the
proportion of cells $F=i/n$ and on the y-axis the corresponding
proportion of commuters $L$ with:
\begin{equation}
L(i)=\frac{\sum_{j=1}^iC_R^j}{\sum_{j=1}^nC_R^j}
\end{equation}
If all the cells had the same number of residents/workers the Lorenz
curves would be the diagonal from $(0,0)$ to $(1,1)$. In general we
observe a concave curve with a more or less strong curvature. In the
Lorenz curve, the stronger the curvature the stronger the inequality
and, intuitively, the smaller the number of hotspots. This remark
allows us to construct a criterion by relating the number of dominant
places (i.e. those that have a very high number of residents/workers
compared to the other cells) to the slope of the Lorenz curve at point
$F=1$: the larger the slope, the smaller the number of dominant values
in the statistical distribution. The natural way to identify the
typical scale of the number of hotspots is to take the intersection
point $F^*$ between the tangent of $L(F)$ at point $F=1$ and the
horizontal axis $L=0$ (see Suplementary Figure S3). This method is inspired from the classical
scale determination for an exponential decay: if the decay from $F=1$
were an exponential of the form $\exp{-(1-F)/a}$ where $a$ is the
typical scale we want to extract, this method would give $1-F^*=a$
(see \cite{Louail:2014} for further discussion).

\subsubsection*{Supplementary note 4}
%Sensitivity of the ICDR values to the density threshold  considered to define hotspots}
\label{sec:ICDR-sens}

On Supplementary Figure S8 we plot the
$I$,$C$,$D$,$R$ values of the 31 Spanish urban areas as a function of the
density threshold $\rho^*$ chosen to define hotspots (here defined
relatively to the density value $\rho_{LB}$ returned by the 'LouBar'
method -- see section \ref{sec:hotspots-def})).

In the extreme case represented on Supplementary Figure S9 all cells whose number of residents/workers
are greater than the mean value of the distribution of
residents/workers are tagged as hotspots (see \cite{Louail:2014} for a
discussion of this criteria and his comparison to the "LouBar''
criteria used in this study). With a much broader acceptation of what
is an hotspot, it is obvious that the term $I$ will increase
drastically since we increase both the number of residential hotspots
and the number of work hotspots. Still what is important to notice is
that we can still observe the qualitative trend observed with the
LouBar method. As the population size increases, the decrease of the
proportion $I$ of integrated flows is accompanied by an increase of
the proportion $R$ of random flows.

\subsubsection*{Supplementary note 5}
%Null model} 
\label{sec:null-model}

In order to evaluate to what extent the $ICDR$ values of a given city
are characteristic of its commuting structure, we compare these values
to the values returned by a reasonable null model of commuting
flows. The guiding idea is to generate OD matrices that (i) have the
same size than the city's OD matrix (i.e that contain the same total
number of individuals) ; (ii) that respect the city's static spatial
organisation (i.e. the in- and out-degrees of all nodes should stay
constant) ; and (iii) that randomize the flows between the nodes,
i.e. with different weights of the edges. Such a null model of flows
that respects the static organization of the city is indeed more
reasonnable and realistic than a null model that would respect the
total number of individuals in the matrix but that would modify the
in- and out-degrees of the nodes.

To generate a random graph that conserves the in- and out- degree of
each node of the reference graph, we use the Molloy-Reed algorithm
\cite{Molloy:1995} which complexity is in $O(n)$, where $n$ is the sum
of the weights of the edges (i.e. the number of individuals in the OD
case).

\subsubsection*{Supplementary note 6}
%Sensitivity of the classification of cities to the density
%  threshold considered to define hotspot}
\label{sec:clusters-sens}

In order to evaluate the sensitivity of the classification of cities
to the number of hotspots selected in cities, for each city $i$ we
make vary the number of work hotspots between $H_w^{i}$ the reference
value returned by the LouBar method (see Supplementary Note 3),
%\ref{sec:hotspots-def})
 and two times the reference value $2\times
H_w^{i}$: $H_w^{LB}(1 + \delta), \delta \in [0;1]$. As for the
sensitivity to noise in the flows $C_{ij}$, we evaluate the stability
of the classification of cities in $k$ groups with the Jaccard index
(see Supplementary Note \ref{sec:ICDR-sens}). The values of $J_I$ as a
function of $\delta$ are represented on Supplementary
Figure S13.

%\begin{thebibliography}{1}
%\bibitem{Louail:2014} Louail, T. et al. From mobile phone data to the
%  spatial structure of cities. \emph{Scientific
%    reports}~\textbf{4}:5276 (2014).
%\bibitem{Molloy:1995} Molloy, M. \& Reed, B. A critical point for
%  random graphs with a given degree sequence. \emph{Random structures
%    \& algorithms}, \textbf{6}(2‐3), 161--180 (1995).
%\end{thebibliography}

\end{document}